\documentclass{aa}
\usepackage[utf8]{inputenc}
\usepackage[flushleft]{threeparttable}
\usepackage{lscape}
\usepackage{floatrow}
\usepackage[utf8]{inputenc}
\long\def\/*#1*/{}

\title{X-ray spectra of the Fe-L complex}
\author{Liyi Gu \inst{1,2}
\and 
A.J.J. Raassen \inst{2,3}
\and
Junjie Mao \inst{4,2}
\and 
Jelle de Plaa \inst{2}
\and 
Chintan Shah \inst{5}
\and 
Ciro Pinto \inst{6}
\and 
Norbert Werner \inst{7,8,9}
\and 
Aurora Simionescu \inst{2,10,11}
\and 
François Mernier \inst{7,12,2}
\and 
Jelle S. Kaastra \inst{2,10} 
}

\institute{
RIKEN High Energy Astrophysics Laboratory, 2-1 Hirosawa, Wako, Saitama 351-0198, Japan
\and
SRON Netherlands Institute for Space Research, Sorbonnelaan 2, 3584 CA Utrecht, the Netherlands 
\and 
Astronomical Institute ``Anton Pannekoek'', Science Park 904, 1098 XH Amsterdam, University of Amsterdam, The Netherlands
\and
Department of Physics, University of Strathclyde, Glasgow, G4 0NG, UK
\and 
Max-Planck-Institut f$\rm \ddot{u}$r Kernphysik, Heidelberg, D-69117 Heidelberg, Germany
\and 
Institute of Astronomy, Madingley Road, CB3 0HA Cambridge, United Kingdom
\and 
MTA-E$\rm \ddot{o}$tv$\rm \ddot{o}$s University Lend$\rm \ddot{u}$let Hot Universe Research Group, P$\rm \acute{a}$zm$\rm \acute{a}$ny P$\rm \acute{e}$ter 
s$\rm \acute{e}$t$\rm \acute{a}$ny 1/A, Budapest, 1117, Hungary
\and 
Department of Theoretical Physics and Astrophysics, Faculty of Science, Masaryk University, Kotl$\rm \acute{a}$$\rm \breve{r}$sk$\rm \acute{a}$ 2, Brno, 611 37, Czech Republic
\and 
School of Science, Hiroshima University, 1-3-1 Kagamiyama, Higashi-Hiroshima 739-8526, Japan
\and
Leiden Observatory, Leiden University, PO Box 9513, 2300 RA Leiden, the Netherlands 
\and
Kavli Institute for the Physics and Mathematics of the Universe (WPI), University of Tokyo, Kashiwa 277-8583, Japan  
\and 
Institute of Physics, E$\rm \ddot{o}$tv$\rm \ddot{o}$s University, P$\rm \acute{a}$zm$\rm \acute{a}$ny P$\rm \acute{e}$ter 
s$\rm \acute{e}$t$\rm \acute{a}$ny 1/A, Budapest, 1117, Hungary
}

\date{November 2018}

\abstract{

The {\it Hitomi} results on the Perseus cluster lead to improvements in our knowledge of atomic physics which are crucial for the precise 
diagnostic of hot astrophysical plasma observed with high-resolution X-ray spectrometers. However, modeling uncertainties 
remain, both within but especially beyond {\it Hitomi}'s spectral window. A major
challenge in spectral modeling is the Fe-L spectrum, which is basically a complex assembly of $n\geq3$ to
$n=2$ transitions of Fe ions in different ionization states, affected by a range of atomic processes
such as collisional excitation, resonant excitation, radiative recombination, dielectronic recombination,
and innershell ionization. In this paper we perform a large-scale theoretical calculation on each of the 
processes with the flexible atomic code (FAC), focusing on ions of \ion{Fe}{XVII} to \ion{Fe}{XXIV} that form
the main body of the Fe-L complex. The calculation includes a large set of energy levels with a broad range of  
quantum number $n$ and $l$, taking into account the full-order configuration interaction and all possible resonant
channels between two neighbour ions. The new data are found to be consistent within 20\% with the recent individual
$R$-matrix calculations for the main Fe-L lines, although the discrepancies become significantly larger for the 
weaker transitions, in particular for \ion{Fe}{XVIII}, \ion{Fe}{XIX}, and \ion{Fe}{XX}. By further testing the new FAC calculations with 
the high-quality RGS data from 15 elliptical galaxies and galaxy clusters, we note that the new model 
gives systematically better fits than the current SPEX v3.04 code, and the mean Fe abundance
decreases by 12\%, while the O/Fe ratio increases by 16\% compared with the results from the current code. Comparing the FAC fit results
to those with the $R$-matrix calculations, we find a temperature-dependent discrepancy of up to $\sim 10$\% on the Fe abundance between
the two theoretical models. 
Further dedicated tests with both observed spectra and targeted laboratory measurements
are needed to resolve the discrepancies, and ultimately, to get the atomic data ready for the next high-resolution X-ray spectroscopy
mission. }

\titlerunning{Fe-L spectrum }
\authorrunning{L. Gu}

\begin{document}

\maketitle

\section{Introduction}
\label{sec:intro}

Great and persistent efforts have been spent on modeling the collisionally-ionized hot plasma for astrophysical diagnostics \citep{cox1969, landini1972, mewe1972, ray1977, smith2001}.
Several computer codes have been developed in order to explain the observed X-ray emission and to understand the underlying physics
of objects.
Major improvements in the plasma modeling codes were driven by the ever-increasing sensitivity and spectral resolution of X-ray instruments. 
The early plasma models, including only the strongest emission lines from each ion, were sufficient to fit 
most of the spectra obtained with the proportional counters on the {\it Einstein}, {\it EXOSAT}, and {\it ROSAT} missions (spectral 
resolution $R < 10$, e.g., \citealt{jones1984}). The X-ray CCDs on {\it ASCA}, {\it Chandra}, and {\it XMM-Newton}
can better resolve the spectrum with $R$ of $10-60$, motivating the updates on the plasma codes to include better
calculations of the detailed ionization balance and satellite line emission (e.g., \citealt{kaastra1992}). These calculations
were found still inadequate for explaining the fully-resolved spectra ($R = 50-1300$) obtained with the grating instruments onboard
{\it Chandra} and {\it XMM-Newton}, and most recently, the micro-calorimeter experiment on the {\it Hitomi} satellite. Over time,
previous calculations of collisional plasma have evolved into the three main codes: AtomDB/APEC \citep{smith2001, foster2012}, SPEX \citep{kaastra1996},
and Chianti \citep{dere1997, dz2015}. 

Plasma models are built on a substantial database of atomic structure and reaction rates, which can only be completed using theoretical calculations.
Only a few key parameters have been verified against laboratory measurements. The unavoidable uncertainties in the theoretical results
have propagated into a significant budget of errors in the astrophysical measurements, giving challenges to the scientific interpretation. 
As reported in \citet{atomic2017}, the {\it Hitomi} observation of the Perseus cluster provides a textbook example showing the challenges: 
the difference between the APEC and SPEX measurements of the Fe abundance is 16\%, which is 17 times higher than the statistical uncertainty,
and 8 times higher than the instrumental calibration error. The discrepancies between the two codes are mostly on detailed collisional excitation
and dielectronic recombination rates of \ion{Fe}{XXIII} to \ion{Fe}{XXVI} ions. It becomes clear that high-resolution X-ray spectroscopy is 
now heavily relying on the plasma modeling and the underlying atomic data.

It should be noted that the {\it Hitomi} data can only test K-shell atomic data in the $2-10$~keV band due to the closed gate valve. 
The model uncertainties of the X-ray band beyond {\it Hitomi}'s spectral window, in particular for the Fe-L complex, remain mostly unknown. 
Substantial work is clearly needed to verify these bands before the launch of the next {\it Hitomi}-level mission. 

%The initial study of the {\it Hitomi} data \citep{atomic2017} also revealed a number of places where the current atomic models needed to update. Measurements with 
%the pre-launch versions of APEC and SPEX differed by 8\% and 35\% in temperature and Fe abundance, respectively. Many of the disagreements were
%addressed by updating wavelengths and cross sections, as well as by fixing bugs. As a result, two codes agree better, though not fully 
%consistent, upon the spectrum
%from a 4~keV hot plasma in the $2-10$~keV band. Substantial work remains to verify the model calculation of the other X-ray bands (e.g., Fe-L), 
%and/or for emission from other astronomical sources.

The Fe-L emission from \ion{Fe}{XVII} to \ion{Fe}{XXIV} is observed from astrophysical bodies as diverse as the solar flare/corona, interstellar
medium, supernova remnants, and galaxy clusters. The Fe-L lines are often very bright, frequently used as diagnostics of electron temperature (e.g.,
\citealt{smith1985}), electron density (e.g., \citealt{p1996}), and chemical abundances \citep{werner2006, dp2017}. The large oscillator strength 
of some Fe-L resonance lines, for instance, the \ion{Fe}{XVII} $2p-3d$ transition at 15~{\AA} and the \ion{Fe}{XVIII} $2p-3d$ transition at 14.2~{\AA}, provide
a unique opportunity for observing resonance scattering in stellar coronae and galaxy clusters \citep{gil1987, xu2002}. The resonance scattering
is one of the few available tools to determine the isotropic gas motion in the hot plasma \citep{churazov2010, gu2018b}.

The rich science of Fe-L motivated a number of theoretical efforts on the spectral modeling, in particular for \ion{Fe}{XVII}. Based on the early
distorted-wave scattering calculations, \citet{smith1985, gold1989, chen1989} reported that the indirect excitation, e.g., the resonant excitation, 
has a significant contribution to the some of the Fe-L lines. \citet{feldman1995} pointed out that the innershell ionization of \ion{Fe}{XVI} might
be another channel to excite \ion{Fe}{XVII}. However, even though various effects were taken into account in these models, they still showed significant 
discrepancies with observations. The spectrum of the solar corona, obtained with the Solar Maximum Mission flat crystal spectrometer, showed that
the early models significantly overestimated the \ion{Fe}{XVII} $2p-3d$ line at 15~{\AA} \citep{p1996}, and the intensity ratio of this line to 
its neighbour intercombination line at 15.26~{\AA}, often labeled $I_{\rm 3C}/I_{\rm 3D}$,
was consistently lower than the calculations. Ground experiments using the electron beam ion trap and other devices indicated a similar bias 
\citep{brown1998, bernitt2012, shah2019}. As a related issue, the {\it Chandra} and {\it XMM-Newton} grating
observations of stellar coronae produced a range of \ion{Fe}{XVII} $2p-3s$/$2p-3d$ ratios \citep{brinkman2000, audard2001}, which were not fully consistent
with the values from early theoretical models. The same discrepancies were seen in elliptical galaxies \citep{xu2002} and supernova remnants \citep{behar2001}.

The tension between the early theory and observation on the Fe-L has been partially lifted by the advent of follow-up calculations.  
Based on an improved distorted wave calculation, \citet{gu2003} (hereafter G03) revisited the direct and indirect line formation processes of Fe-L. 
G03 also improved the collisional-radiative modeling, allowing a more accurate calculation of the cascading contribution to the main spectral line
intensities. Fits using the G03 model to the {\it XMM-Newton} and {\it Chandra} grating spectra of Capella showed a reasonable agreement \citep{gu2006}.
Recently, $R$-matrix scattering calculations have been performed for \ion{Fe}{XVII} by 
\citet{aggarwal2003}, \citet{chen2002}, \citet{loch2006}, and \citet{liang2010}, as well as for other Fe-L species (\citealt{witt2006} for \ion{Fe}{XVIII}, 
\citealt{butler2008} for \ion{Fe}{XIX}, \citealt{witt2007} for \ion{Fe}{XX}, \citealt{badnell2001} for \ion{Fe}{XXI}, 
\citealt{liang2012} for \ion{Fe}{XXII}, \citealt{fern2014} for  \ion{Fe}{XXIII}, and \citealt{liang2011} for \ion{Fe}{XXIV}). Benchmarks with observational/laboratory 
data using the $R$-matrix results showed significant improvements over the early distorted-wave models for individual ions 
\citep{del2005, del2006, del2006c, del2011}.  Both the G03 and $R$-matrix models are now commonly used in astrophysics, 
although it is found that some discrepancies might still exist between the two calculations \citep{butler2008, brown2008, liang2011, del2011, aggarwal2013}.

In this paper, we present a new systematic calculation of the Fe-L spectrum for optically-thin collisionally-ionized plasma. The calculation is 
based on the atomic structure and distorted wave scattering calculation by the FAC atomic code, and the line formation calculation by the SPEX plasma code. 
We aim to perform a consistent large-scale calculation of the fundamental data for all the Fe-L
species (\ion{Fe}{XVII} to \ion{Fe}{XXIV}), focusing mainly on the dominant indirect excitation processes: the resonant excitation and dielectronic recombination.
Compared to G03, our work adopts the updated FAC code, expands the intermediate states of the indirect processes, and calculates up to higher excited levels 
(see \S~\ref{sec:vg03} for details). The new results are compared systematically to the previous theoretical calculations, and are tested using the 
observational data obtained with the {\it XMM-Newton} grating spectrometer.

A systematic (re-)calculation of the Fe-L complex is useful in the following two aspects. First, the comparison of models from the latest distorted wave
code with those from the available $R$-matrix works will potentially allow us to identify problem areas where discrepancies still occur among
the theories. Such information will be useful for experimentalists to set priority on the laboratory astrophysical measurements needed to benchmark the theoretical calculation. 
Second, fitting the astrophysical spectra with the new and the available calculations would show the variations of source parameters caused by the underlying atomic database. 
Potentially, one might take such variations into account as one of the systematic uncertainties on the measurements, which might affect the scientific interpretation of the 
observed data.

Structure of the paper is present as follows. Section~\ref{sec:method}
describes the theoretical approach. Section~\ref{sec:result} presents the results and the comparison with other theoretical data. Section~\ref{sec:apply} discusses
the impact of the new calculation on the existing high-resolution astrophysical measurements.

\section{Theoretical method}
\label{sec:method}

Astrophysical plasmas in diffuse objects is often found in collisional ionization equilibrium (CIE), usually characterized by 
low density (e.g., $10^{-4}-10^{-1}$ cm$^{-3}$ in galaxy clusters). Albeit of low collisional frequency, the electron impact
excitation, followed by radiative cascade, is often the key process to produce X-ray line emissions from ions. The direct electron-ion collision cross sections for highly charged ions can be calculated by common theoretical tools based on Coulomb-Born and distorted-wave approximations \citep{mewe19722}. However, these tools cannot tackle at once the indirect contributions, such as autoionizing resonances, dielectronic recombination and innershell ionization. We focus below a manual calculation of the indirect excitations, mainly for the ionic species producing the Fe-L lines. The rates coefficients of the direct excitation are also calculated for the relevant levels.

\begin{figure}[!htbp]
\resizebox{0.9\hsize}{!}{\includegraphics[angle=0]{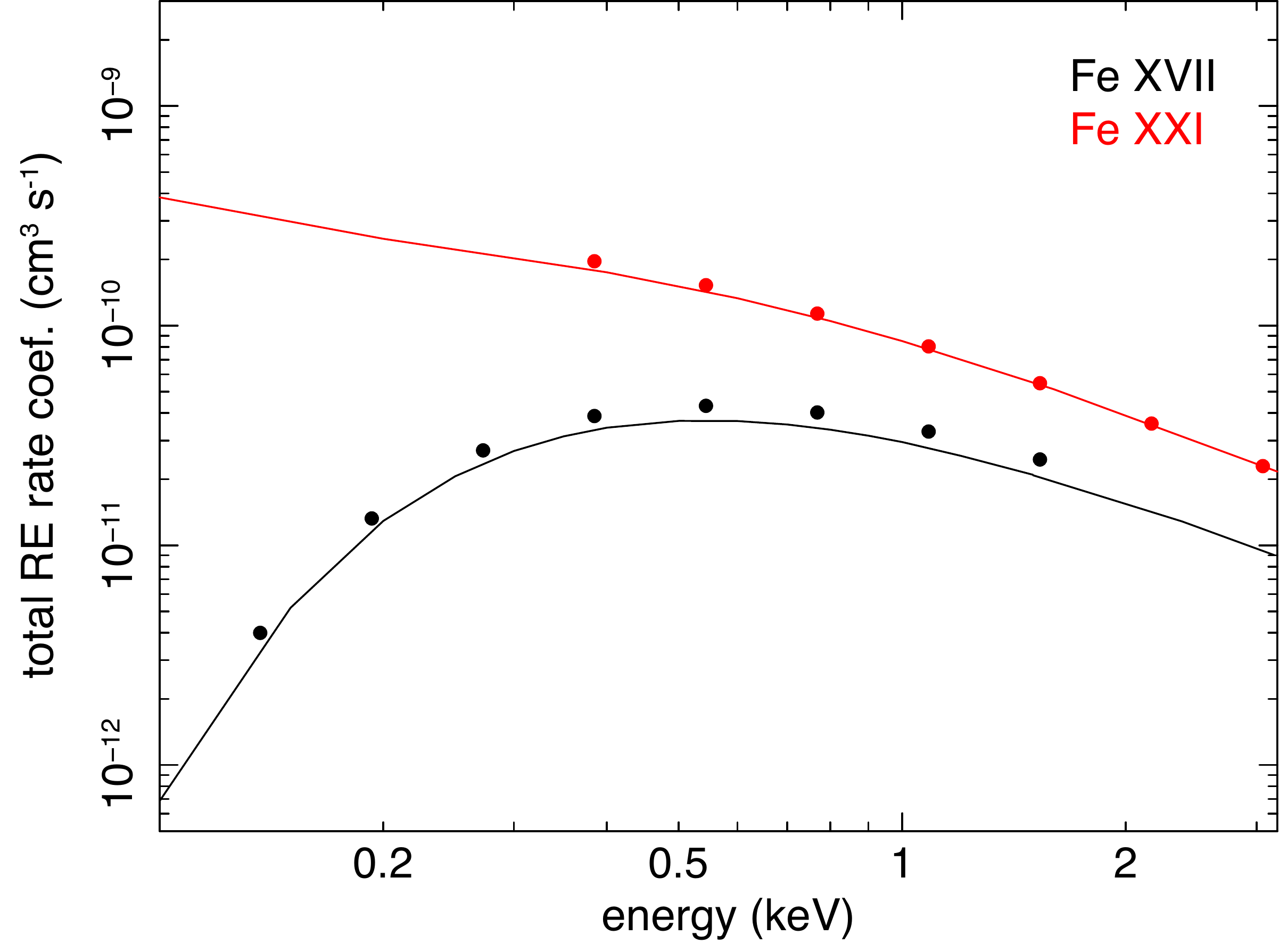}}
\caption{Total resonant excitation (RE) rate coefficients of \ion{Fe}{XVII} and \ion{Fe}{XXI} through their neighbor ions as a function of energy. The
ground states are not included. The solid lines show the present work, and the data points are taken from \citet{gu2003}. }
\label{fig:totre}
\end{figure}

\subsection{Resonant excitation}

Resonant excitation can be understood as a two-step process. First a free electron is captured by the target ion, with the accompanying excitation of a bound electron, giving a 
doubly excited level 
in the lower ionization state. The doubly excited level will then decay by radiation or Auger process. The Auger decay to an excited level of the initial ionization 
state will effectively contribute to the excitation of the target ion. 

We calculate the resonant excitation from an initial state $i$ to the final state $f$, via a doubly excited state $d$. Both states $i$ and $f$ have ionic charge $q$, and 
the state $d$ has a charge $q-1$.
Assuming a thermal plasma, the dielectronic recombination rates are calculated from the inverse process, autoionization, by the detailed balance,
\begin{equation}
R^{DR}_{id} = n_{e} n_{q} \frac{g_{d}}{2g_{i}} A^{a}_{di} \left(\frac{h^{2}}{2 \pi mkT}  \right)^{3/2} e^{- E_{x}/kT}, 
\end{equation}
where $n_{e}$ and $n_{q}$ are the densities of electrons and the target ions, $g_{d}$ and $g_{i}$ are the statistical weights of the intermediate and initial states, $h$ is the Planck constant, $m$ is the mass of the charge, $T$ is the equilibrium temperature, and $A^{a}_{di}$ and $E_{x}$ are the rate and energy of the Auger transition, respectively. The chance of excitation to the final state $f$ is given by the 
branching ratio
\begin{equation}
B^{RE}_{df} = \frac{A^{a}_{df}}{\Sigma (A^{r}_{d} + A^{a}_{d})},
\end{equation}
where $A^{a}_{df}$ is the Auger rate from the intermediate state to the final state, and $A^{r}_{d}$ and $A^{a}_{d}$ are the radiative and Auger transitions pertaining to the state $d$, respectively. Hence, the resonant excitation rate can be calculated as
\begin{equation}
R^{RE}_{if} = \Sigma_{d} R^{DR}_{id} B^{RE}_{df}.
\end{equation}

The atomic structure of the initial, intermediate, and final states, as well as the related transitions, are all computed with FAC version 1.1.4 
\citep{gu2008} in a fully relativistic way. The distorted-wave approximation is used for interaction with the continuum states. The relativistic electron-electron interactions (Coulomb + Breit form) in the atomic central potential are considered, while the higher-order electronic interactions,
which are hard to be described by an analytic model, are approximated by the configuration mixing of the bound states.

%For the coronal/nebular ($< 10^{14}$ cm$^{-3}$) plasma, we calculate the resonance only from the ground state. 
For high density plasma, the excitation only from the ground state might not be sufficient. 
As shown in Appendix A, the low-lying metastable levels become significantly populated at density $> 10^{12}$ cm$^{-3}$, 
and the excitation and recombination from these levels are required to produce the model spectrum. 
For each ion, we include three lowest excited levels, as well as the ground, as the initial states $i$.
The three levels are sufficient for modeling the coronal plasma ($< 10^{14}$ cm$^{-3}$), while for a higher
density, more metastable levels at higher energies are then required \citep{badnell2006}.

It is crucial to include a large set of configuration for the autoionizing intermediate 
state $d$, as leaving some states out would cause insufficient resonant excitation and incomplete configuration interaction \citep{badnell1994}. We maximize the configurations for each $n$ group, for instance, for \ion{Fe}{XVII} excitation, the relevant \ion{Fe}{XVI} states 2$s^2$2$p^5$3$lnl'$, 2$s$2$p^6$3$lnl'$ ($3 \leq n \leq 15$),
2$s^2$2$p^5$4$lnl'$, and 2$s$2$p^6$4$lnl'$ ($4 \leq n \leq 15$) are all included in the calculation.  
The singly excited levels, 2$s^2$2$p^6nl'$ ($3 \leq n \leq 15$) are also taken into account for determining the radiative transitions and branching ratios. 
A complete set of quantum numbers $l'$ is included. The \ion{Fe}{XVI} atomic structure then contains $\sim 30000$ states.
For each doubly excited state, the radiative cascade rates to the lower bound states and the autoionization rates back to \ion{Fe}{XVII} states are computed to derive the detailed branching ratios. Radiative transitions of electric dipole (E1), electric quadrupole (E2), magnetic dipole (M1),
and magnetic quadrupole (M2) types are considered for the cascades. The numbers of radiative transitions are 
$\sim 200000 - 300000$ for a typical group of $d$ states with the same principle quantum number for \ion{Fe}{XVI}. The number
increases to more than 500000 for \ion{Fe}{XVIII} $-$ \ion{Fe}{XX}.

The calculation considers the radiative cascades of the $d$ states followed by autoionization. For instance, the 2$s$2$p^6$3$lnl'$ might turn into 2$s^2$2$p^5$3$lnl'$ through
a $2p-2s$ transition, and then autoionize to \ion{Fe}{XVII}. This consists of a multi-step resonance. In principle, the radiative cascade should be traced down to the ground, while
practically the strength of the resonance decays quickly by the branching ratio at each step, and the contribution can be ignored after two steps of cascades.

We include a sufficient amount of final states for the autoionization. For \ion{Fe}{XVII}, the final configurations are 2$s^2$2$p^6$, 
2$s^2$2$p^5$3$l$, 2$s$2$p^6$3$l$, and 2$s^2$2$p^5$4$l$. The Auger rates from all the $d$ states to the $f$ states are calculated.
The numbers of Auger transitions vary from $\sim 1000 - 40000$ for different groups of $n-$resolved intermediate states. In some cases when the 
bound electron is highly excited after autoionization (e.g., 
for some of the 2$s$2$p^6$4$lnl'$ channels), we calculate the radiative cascades down to the selected final configurations.

The contributions from high Rydberg states are taken into account by extrapolation. In the \ion{Fe}{XVII} case, the resonances via 
2$s^2$2$p^5$3$lnl'$ and 2$s^2$2$p^5$4$lnl'$ ($16 \leq n \leq 100$) are calculated by a $n^{-3}$ scaling on the Auger rates 
based on the results from the lower states. As shown in \S\ref{sec:res_result}, the actual $n$-dependence appears to scatter around
the assumed scaling, which would bring an uncertainties of $\leq 3$\% 
to the total resonance strength.

\begin{figure*}[!htbp]
\centering
\resizebox{\hsize}{!}{\includegraphics[angle=0]{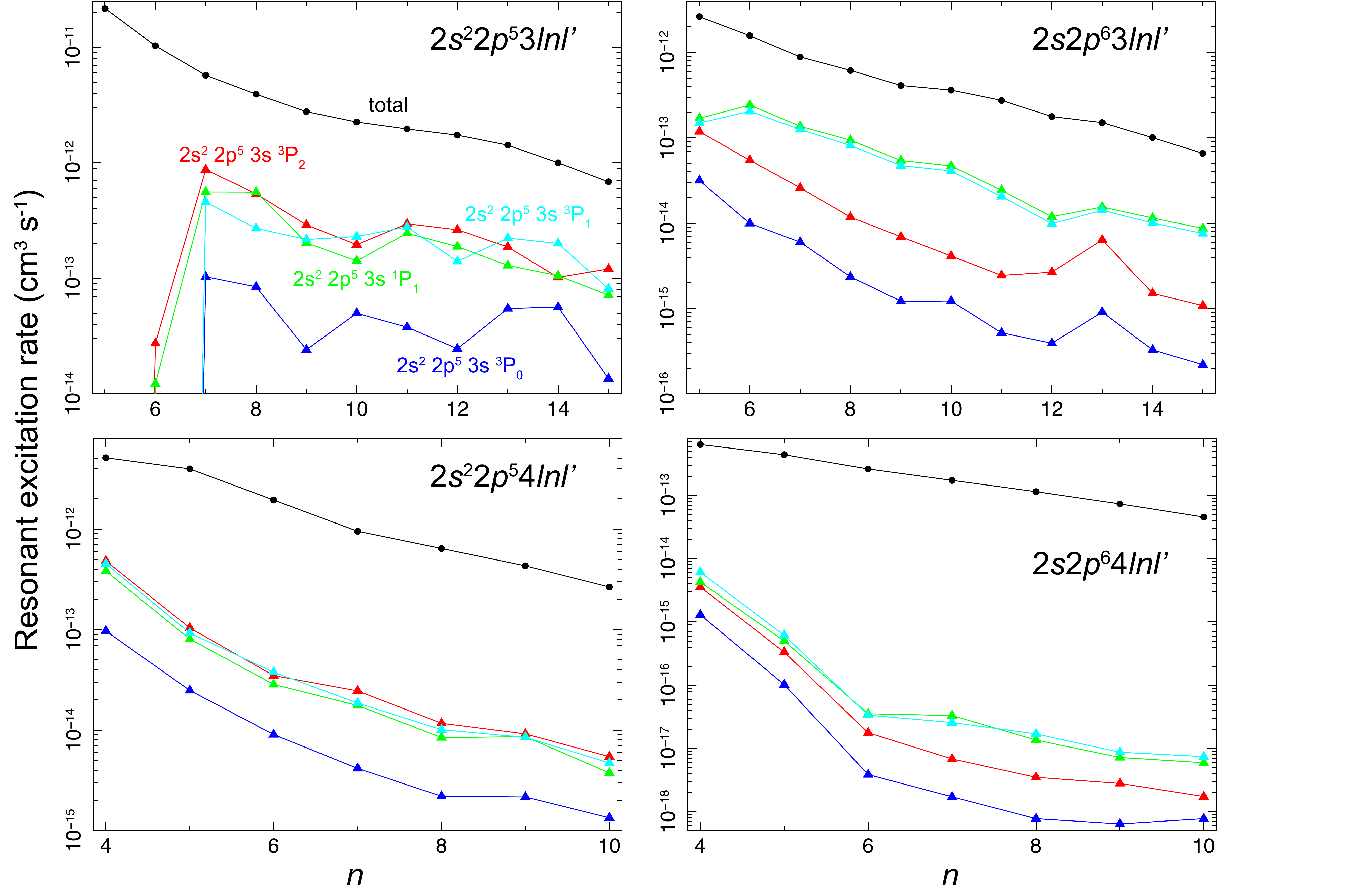}}
\caption{Resonant excitation rate coefficients for \ion{Fe}{XVII} at an energy of 0.4~keV as a function of principle quantum number
$n$. The four panels plot the resonances through four main autoionizing \ion{Fe}{XVI} states: 2$s^2$2$p^5$3$lnl'$, 2$s$2$p^6$3$lnl'$, 2$s^2$2$p^5$4$lnl'$, and 2$s$2$p^6$4$lnl'$. 
The autoionization into the lowest four excited states of \ion{Fe}{XVII} are highlighted with four different colors. }
\label{fig:ndep}
\end{figure*}

\begin{figure}[!htbp]
\resizebox{0.9\hsize}{!}{\includegraphics[angle=0]{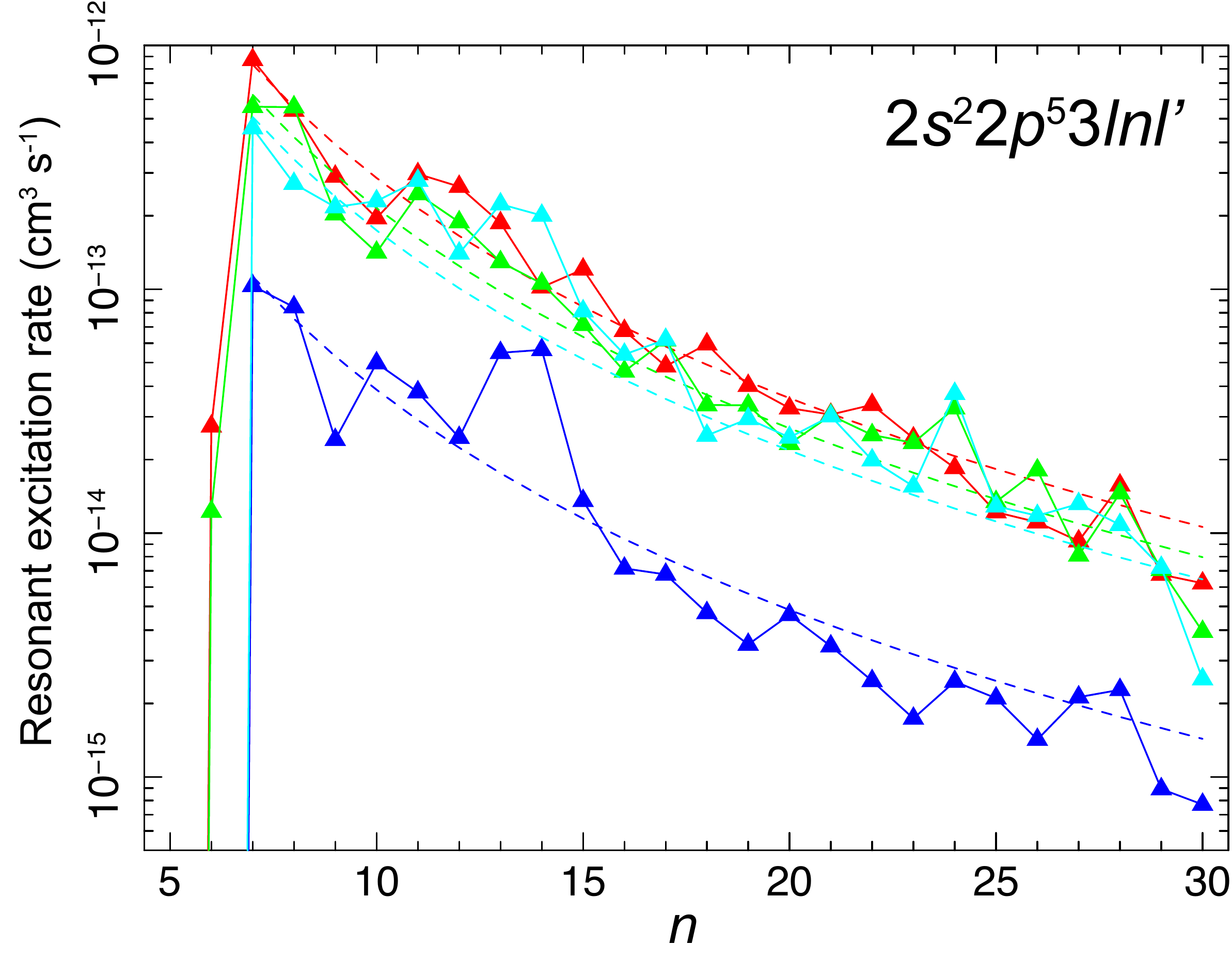}}
\caption{Resonant excitation rate coefficient of the lowest four excited states of \ion{Fe}{XVII} through \ion{Fe}{XVI} 
states 2$s^2$2$p^5$3$lnl'$ at 0.4~keV, as a function of principle quantum number $n$ up to 30. The dashed lines show the 
scaling functions for estimating the contribution from the high-$n$ states. }
\label{fig:nscaling}
\end{figure}

\subsection{Dielectronic recombination}
\label{sec:method_dr}

Dielectronic recombination is one of the most dominant channels of indirect excitation. The DR itself is very similar to electron-impacting excitation,
except that the final state of the impact electron is in a bound state rather than in the continuum. Many of the excitation channels via DR are already
incorporated in the current SPEX database, version 3.04. However, a few of them are still missing. To update the atomic database, we carry out a new 
calculation for a complete set of DR capture channels using the FAC code.

%Here, we calculate these DR capture channels as a complement to the standard SPEX code.

We consider DR from an initial state $i$ to a final state $f$, via a doubly excited state $d$. While for the resonances states $i$ and $f$ have the same charge,
here $f$ has a charge $q$ and $i$ has $q+1$. The DR rates can be obtained as
\begin{equation}
R^{DR}_{if} = \frac{n_{q+1}}{n_{q}} R^{DR}_{id} B^{DR}_{df},
\label{eq:dr}
\end{equation}
where $n_{q+1}/n_{q}$ is obtained from the ionization balance between ions $q+1$ and $q$, and 
\begin{equation}
B^{DR}_{df} = \frac{A^{r}_{df}}{\Sigma (A^{r}_{d} + A^{a}_{d})}.
\end{equation}
We adopt the new ionization concentration presented in \citet{u17}, which updated the rate equations for the direct collisional ionization and excitation-autoionization.

The DR rates are calculated in a similar way as the resonant excitation process. We set the initial state to the ground, and include a large sets of intermediate states.
For \ion{Fe}{XVII}, the configurations 2$s^2$2$p^4$3$lnl'$, 2$s$2$p^5$3$lnl'$ ($3 \leq n \leq 7$, $l' \leq 5$), 2$s^2$2$p^4$4$lnl'$, and 2$s$2$p^5$4$lnl'$ 
($4 \leq n \leq 7$, $l' \leq 5$) are included in the model. These levels contain a $n=2$ to $n=3$ and $n=4$ excitation of the core electron, associated with an electron 
captured to higher $n$. Although the DR rates for configurations with a $n=1$ to $n=2$, or $n=2$ to $n=2$ core excitation, such as 2$s$2$p^6nl'$ and 1$s$2$s^2$2$p^6nl'$ ($3 \leq n \leq 10$), are already incorporated in the current SPEX database, it is still necessary to include these levels 
in our model to build up a complete cascading network. The same holds for the singly
excited levels 2$s$2$p^5nl'$ ($3 \leq n \leq 10$). Therefore the total levels add up to $\sim 25000$ for \ion{Fe}{XVII}, and more than $30000$ for
\ion{Fe}{XIX} and \ion{Fe}{XX}.

Both the resonant excitation and DR calculations mainly focus on channels through 3$lnl'$ and 4$lnl'$ states. The 3$lnl'$ states are the dominant 
states producing both resonances and DR, depending on the branching ratios of radiative decay and autoionization. The 4$lnl'$ contributes
significantly to the resonant excitation, but much less to the DR.

The stabilization of the doubly excited states by both autoionization and radiative transitions are 
calculated. The radiative cascade is apparently important for the DR calculation,
as initially it populates doubly-excited states with large excitation energies. Practically, we include a small amount of final states of low excitation energies,
and calculate the cascading contributions to these final states corrected for the autoionization loss. For \ion{Fe}{XVII}, the selected final states are 2$s^2$2$p^6$, 
2$s^2$2$p^5$3$l$, 2$s$2$p^6$3$l$, and 2$s^2$2$p^5$4$l$. A full cascading calculation is then done with about 1500000 radiative
transitions, and about 60000 non-radiative transitions. The numbers of transitions increase by a factor of $\sim 5$ for \ion{Fe}{XVIII}
$-$ \ion{Fe}{XX}. 
The further transitions among the final states, and the resulting line power, are calculated with the standard SPEX code. 
In this way we obtain the DR contribution to the main Fe-L lines, while the accompanying satellite lines from the cascade, which often have much longer 
wavelengths and do not affect the Fe-L spectrum, are ignored in this work.

Similar to the resonance calculation, we include the contributions from high Rydberg states (up to $n=100$) by a $n^{-3}$ scaling of the Auger rates.
The extrapolation is done with the cascaded rates for all the selected final states. The scaling is restricted to the dominant DR channels, such as the 3$lnl'$
group in the \ion{Fe}{XVII} case.

\subsection{Innershell ionization}
\label{sec:method_ii}
The innershell collisional ionization of a core electron can enhance the population of excited states \citep{feldman1995}. It depends on two factors: the ionization rate coefficient through electron collisions, and the fractional abundance 
of the neighbour ion with a lower charge state. For \ion{Fe}{XVII}, the effect of the innershell ionization is expected to be small, as the ionization rate is rather small at low temperatures,
and the \ion{Fe}{XVI} to \ion{Fe}{XVII} ratio drops off at high temperatures. As reported in \citet{doron2002} and \citet{gu2003}, the 2$p$ innershell ionization could 
affect the \ion{Fe}{XVII} lines $2p-3s$ transition by $\sim 2 - 3$\%. 

To taken this minor process into account, we apply the innershell ionization rate coefficient data from \citet{gu2003}, which were calculated using the same FAC atomic tool. It includes the ionization of 
both $n=1$ and $n=2$ electrons from the ground. The fractional abundance is calculated based on the new ionization balance of \citet{u17}.

\section{Results}
\label{sec:result}

\subsection{Resonant excitation}
\label{sec:res_result}

The resonant electron-impact excitation rate coefficients are calculated for Fe ions from \ion{Fe}{XVII} to \ion{Fe}{XXV}. 
Fig.~\ref{fig:totre} shows the total resonant excitation rates of \ion{Fe}{XVII} and \ion{Fe}{XXI} as a function of \text{energy}.
They are found to agree with the results from \citet{gu2003} within 10\%. As the theoretical approach of \citet{gu2003} is essentially the same as this work, 
the small discrepancy on the total resonance of \ion{Fe}{XVII} might be caused by
the difference in the input \ion{Fe}{XVI} levels and the branching ratios.

The current approach enables a level-resolved calculation. In Fig.~\ref{fig:ndep}, we plot the $n-$dependent partial resonances for the different flavour of $d$ and $f$ states of 
\ion{Fe}{XVII} excitation. The four lowest excited states, giving the $M2$ magnetic quadrupole forbidden line ($2s^22p^53s$ $^3$P$_{2}$), 
the $3G$ electronic dipole allowed line ($2s^22p^53s$ $^1$P$_{1}$), the $M1$ magnetic dipole forbidden line ($2s^22p^53s$ $^3$P$_{0}$), 
and the $3F$ spin-forbidden intercombination line ($2s^22p^53s$ $^3$P$_{1}$), are highlighted in the plot. The autoionization from 2$s^2$2$p^5$3$lnl'$ is the dominant channel to populate the excited states directly, while the contribution
from 2$s$2$p^6$4$lnl'$ is nearly negligible. The low Rydberg states ($n \leq 5$) of the doubly excited 2$s^2$2$p^5$3$lnl'$ states can autoionize mostly
to the ground of \ion{Fe}{XVII}, the resonances to excited states thus show a sharp rise at $n=7$, and a mild decrease towards higher Rydberg states. For the other $d$ states, the resonances decrease monotonically as a function of $n$ except for a few minor peaks at high-$n$.

As described in Sect.~\ref{sec:method}, the resonance decrease towards high-$n$ is treated by a $n^{-3}$ scaling on the Auger rates fitted to the low-$n$ data.
Previous laboratory measurements of the high-$n$ satellite lines indicated that the actual $n$-dependence sometimes deviates from the theoretical scaling \citep{smith1996}, as the radiative branching ratios would also evolve with the quantum numbers. To assess the uncertainty caused by the $n^{-3}$ assumption, we extend the calculation of $n-$resolved excitation rates from $n=15$ to $n=30$ for the resonances of \ion{Fe}{XVII}. As shown in Fig.~\ref{fig:nscaling}, the actual calculations of the resonance strengths into the four lowly-excited states are compared with the $n^{-3}$ scaling, which is obtained by fitting the excitation rates of $n \leq 15$. Combining the resonances from $n=16$
to $n=30$, the discrepancies between the data and the scaling are $\sim 1-9 \times 10^{-14}$ cm$^{3}$ s$^{-1}$ for the four states. This error
appears to be negligible ($<3$\%) as the total resonant excitation rates are often several $10^{-12}$ cm$^{3}$ s$^{-1}$ for these states.

In Appendix B, we present a systematic comparison of the new calculation with previous results on the excitation rate coefficients of Fe-L. 
The tests, in particular with those from recent $R$-matrix calculations, show agreement within typical errors of $\sim$20\% on the main transitions,
though the discrepancies on the weaker transitions are much larger.
This result agrees with the previous reports (e.g., \citealt{fern2017}). 
Similar conclusions can also be obtained by comparing directly the spectra using the two sets of collisional calculations (\S~\ref{sec:felspectra}).

\begin{figure*}[!htbp]
\centering
\resizebox{0.9\hsize}{!}{\includegraphics[angle=0]{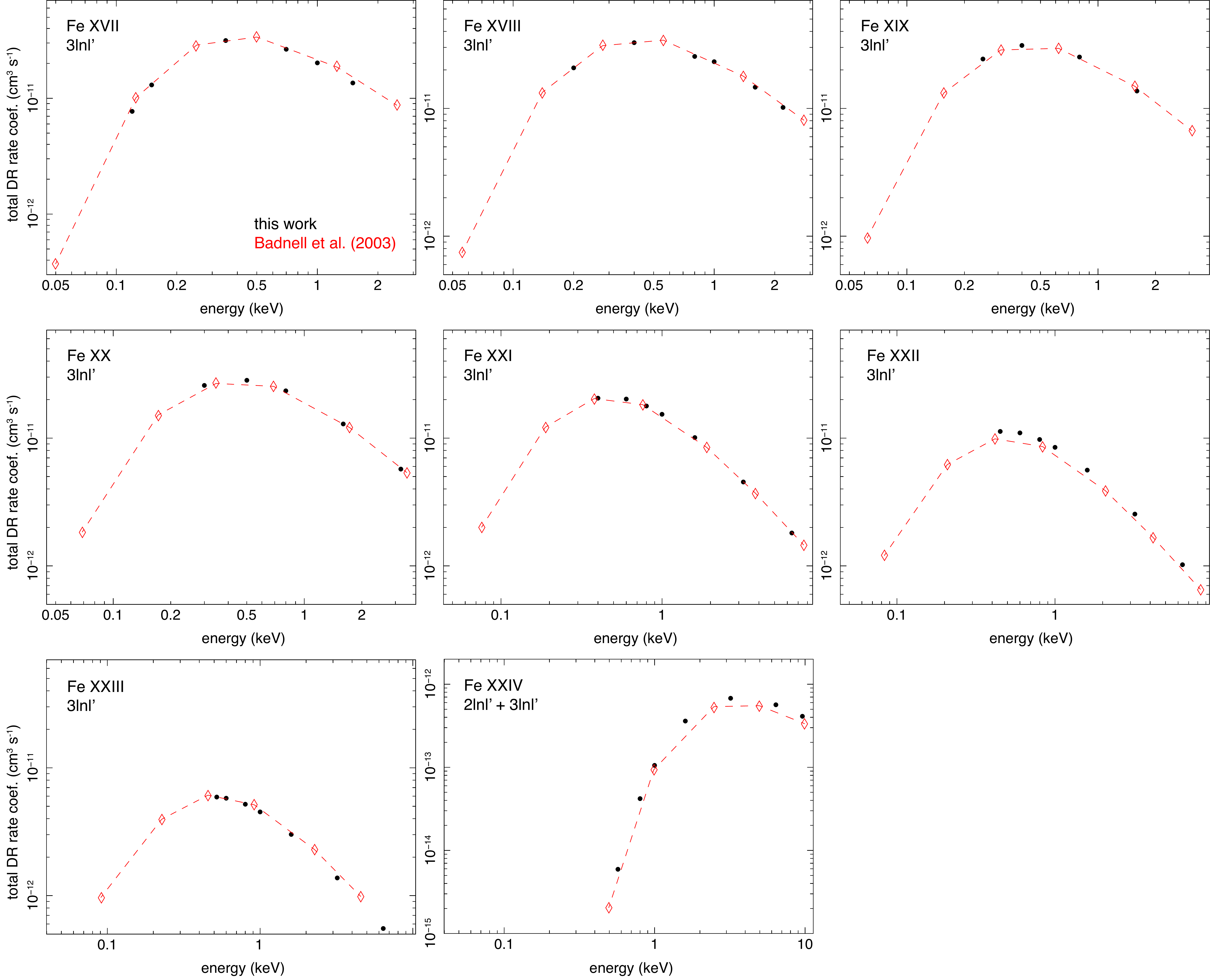}}
\caption{Comparison of total dielectronic recombination (DR) rate coefficients. All plots show captures into 3$lnl'$ states, except
for \ion{Fe}{XXIV} where the combined 2$lnl'$ and 3$lnl'$ are shown. The coefficients include radiative cascades. The large-scale
calculations by \citet{badnell2003} are plotted in red. }
\label{fig:totdr}
\end{figure*}

\subsection{Dielectronic recombination}
\label{sec:dr_result}

As described in Section~\ref{sec:method_dr}, the state-selective dielectronic recombination rates are calculated for each isolated 
channel characterized by the intermediate doubly-excited level $d$. We focus on the $d$ states in which the core electron is excited
from $n=2$ to $n=3$ and 4, and the free electron is captured up to $n = 7$. The 3$lnl'$ channels are much more important than the 4$lnl'$
ones for the DR. Before applying the data in the line formation calculation,
we compare the current results with the state-of-the-art data published by \citet{badnell2003}, which was calculated using 
the Breit-Pauli intermediate coupling approach. 

As shown in Fig.~\ref{fig:totdr}, the two calculations broadly agree upon the total DR rates through 3$lnl'$ to better than 20\%. The 
differences do not appear to be systematic in the energy range for comparison. The main discrepancies are seen in
\ion{Fe}{XXII} at $\sim 0.5$ keV and \ion{Fe}{XXIV} at $> 1$ keV, where our DR rates are higher than the Badnell results by $\sim 15$\%.

%\begin{figure*}[!htbp]
%\centering
%\resizebox{0.9\hsize}{!}{\includegraphics[angle=0]{17A_1234_rate.eps}}
%\caption{Direct contributions to four main \ion{Fe}{XVII} lines, plotted as a function of electron temperature. }
%\label{fig:17arate}
%\end{figure*}

\begin{figure*}[!htbp]
\floatbox[{\capbeside\thisfloatsetup{capbesideposition={right,center},capbesidewidth=6cm}}]{figure}[\FBwidth]
{\caption{Relative contributions to the formation of \ion{Fe}{XVII} 17.09{\AA} (left) and 17.05{\AA} (right) lines, from both excitation and radiative cascades. The source
levels of cascades are plotted in different colors.}\label{fig:17apie}}
{\includegraphics[width=9cm]{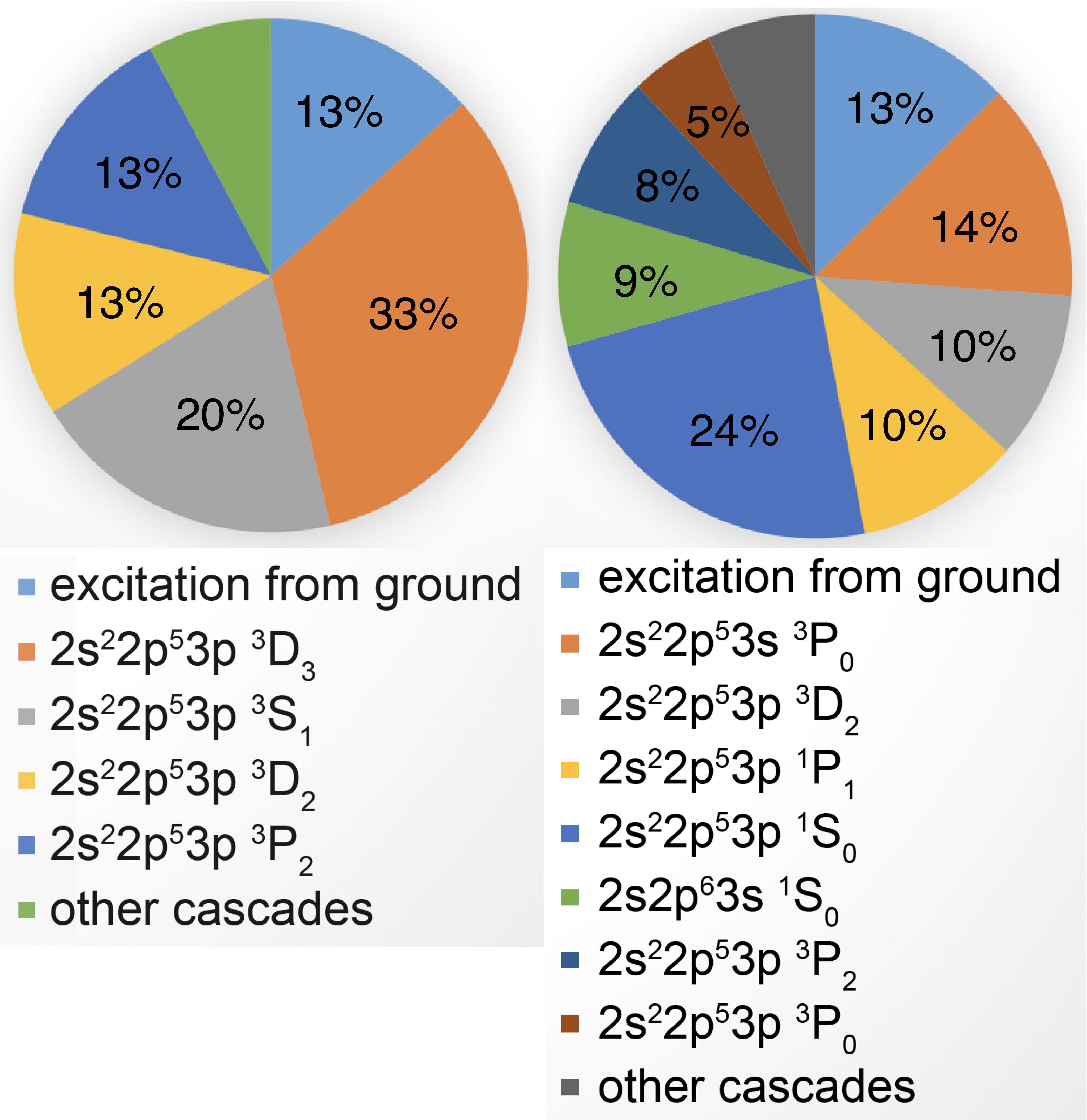}}
%\resizebox{0.5\hsize}{!}{\includegraphics[angle=0]{17A_pie2.eps}}
%\caption{Relative contributions to the formation of \ion{Fe}{XVII} 17.05{\AA} and 17.09{\AA} lines, from both excitation and radiative cascades. The source
%levels of cascades are plotted in different colors.}
\end{figure*}

\begin{figure*}[!htbp]
\centering
\resizebox{0.9\hsize}{!}{\includegraphics[angle=0]{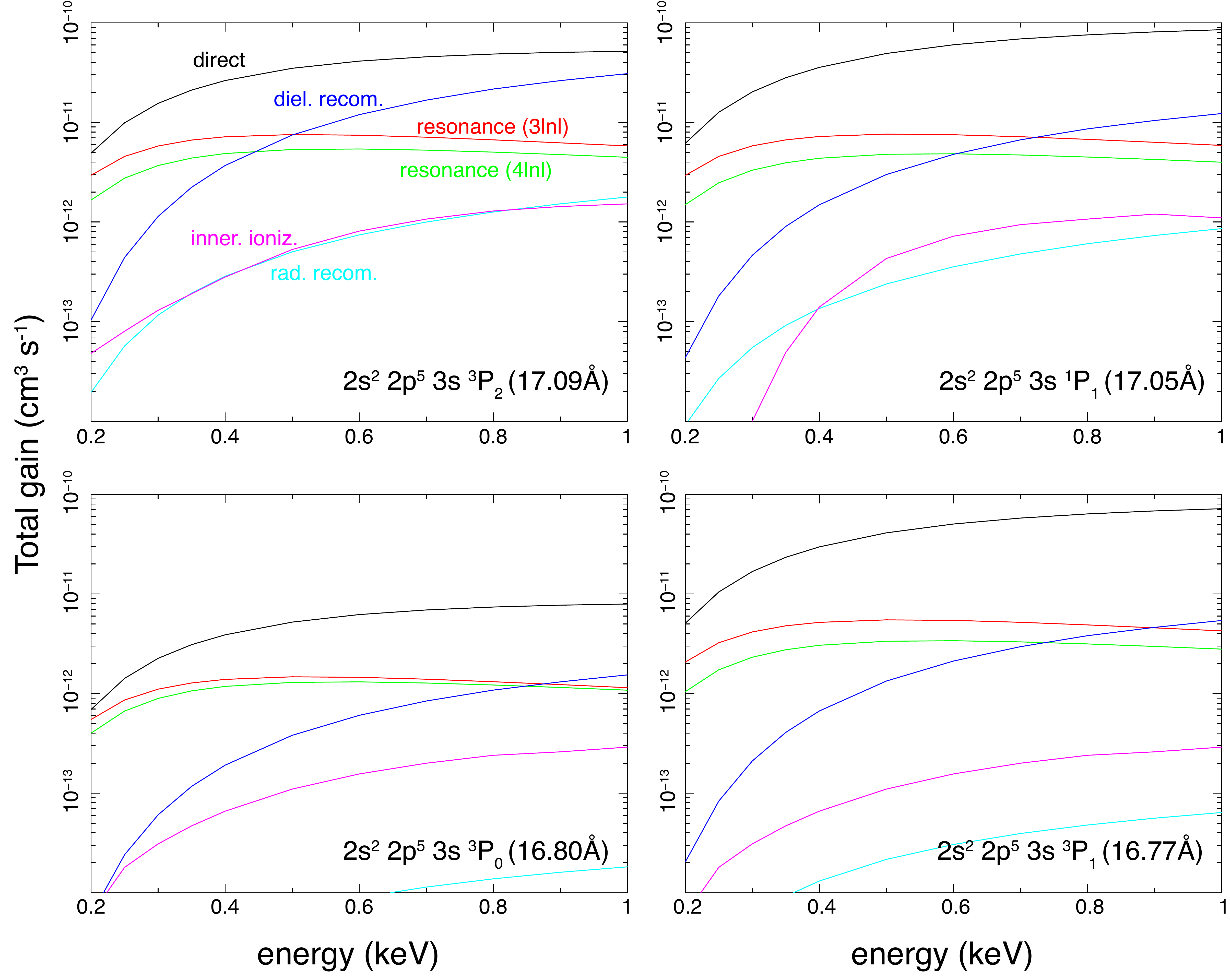}}
\caption{Contributions of rate coefficients to four main \ion{Fe}{XVII} lines after the radiative cascades are taken into account, plotted
as a function of energy. }
\label{fig:17atot}
\end{figure*}

\begin{table*}[!htbp]
\centering
\caption{Levels of the Fe-L ions}
\label{tab:levels}
\begin{threeparttable}
\begin{tabular}{ccccccccccccccccc}
\hline
Index & Z$^{a}$ & Ion$^{b}$ & $n^{c}$ & $L^{d}$ & 2$S$+1$^{e}$ & 2$J^{f}$ & Configuration & Energy (keV)$^{g}$ \\ 
\hline
    0 & 26& 17  & 2   & 0   & 1      & 0    & 2s2.2p6       & 0            \\
    1 & 26& 17  & 3   & 1   & 3      & 4    & 2s2.2p5.3s    & 7.2524E-01   \\
    2 & 26& 17  & 3   & 1   & 1      & 2    & 2s2.2p5.3s    & 7.2714E-01   \\
    3 & 26& 17  & 3   & 1   & 3      & 0    & 2s2.2p5.3s    & 7.3786E-01   \\
    4 & 26& 17  & 3   & 1   & 3      & 2    & 2s2.2p5.3s    & 7.3905E-01   \\
    5 & 26& 17  & 3   & 0   & 3      & 2    & 2s2.2p5.3p    & 7.5549E-01   \\
    6 & 26& 17  & 3   & 2   & 3      & 4    & 2s2.2p5.3p    & 7.5899E-01   \\
    7 & 26& 17  & 3   & 2   & 3      & 6    & 2s2.2p5.3p    & 7.6061E-01   \\
    8 & 26& 17  & 3   & 1   & 1      & 2    & 2s2.2p5.3p    & 7.6174E-01   \\
    9 & 26& 17  & 3   & 1   & 3      & 4    & 2s2.2p5.3p    & 7.6355E-01   \\
   10 & 26& 17  & 3   & 1   & 3      & 0    & 2s2.2p5.3p    & 7.6898E-01   \\
   11 & 26& 17  & 3   & 2   & 3      & 2    & 2s2.2p5.3p    & 7.7106E-01   \\
   12 & 26& 17  & 3   & 1   & 3      & 2    & 2s2.2p5.3p    & 7.7431E-01   \\
   13 & 26& 17  & 3   & 2   & 1      & 4    & 2s2.2p5.3p    & 7.7469E-01   \\
   14 & 26& 17  & 3   & 0   & 1      & 0    & 2s2.2p5.3p    & 7.8772E-01   \\
\hline
\end{tabular}
\begin{tablenotes}
\item[] Note: full-data table can be found via the link to the machine-readable version.
\item[$(a)$] Atomic number.
\item[$(b)$] Isoelectronic sequence number.
\item[$(c)$] Principle quantum number.
\item[$(d)$] Angular momentum quantum number.
\item[$(e)$] Spin quantum number.
\item[$(f)$] Twice the total angular momentum quantum number.
\item[$(g)$] Energies of excited states relative to ground.
\end{tablenotes}
\end{threeparttable}
\end{table*}

\begin{table*}[!htbp]
\centering
\caption{Rate coefficients of the Fe-L in collisional ionization equilibrium}
\label{tab:rate}
\begin{threeparttable}
\begin{tabular}{l@{\:\:}c@{\:\:}ccccccccc}
\hline
Ion & lev$^{a}$ & kT$^{b}$ & CE$^{c}$ & RE$^{d}$ & REc$^{e}$ & RRc$^{f}$ & DRc$^{g}$ & CE+RE$_1^{h}$ & CE+RE$_2^{i}$ & CE+RE$_3^{j}$ \\
\hline
    17 & 1 & 0.1 & 8.960E-15 & 1.227E-13 & 2.548E-13 & 5.777E-17 & 1.927E-17 & 0.00E+00 & 0.00E+00 & 0.00E+00 \\
    17 & 1 & 0.2 & 2.176E-13 & 1.829E-12 & 4.733E-12 & 1.947E-14 & 5.028E-15 & 0.00E+00 & 0.00E+00 & 0.00E+00 \\
    17 & 1 & 0.4 & 8.176E-13 & 4.243E-12 & 1.296E-11 & 2.856E-13 & 3.140E-12 & 0.00E+00 & 0.00E+00 & 0.00E+00 \\
    17 & 1 & 0.8 & 1.151E-12 & 3.855E-12 & 1.321E-11 & 1.260E-12 & 2.227E-11 & 0.00E+00 & 0.00E+00 & 0.00E+00 \\
    17 & 1 & 1.6 & 9.435E-13 & 2.185E-12 & 8.039E-12 & 3.238E-12 & 5.340E-11 & 0.00E+00 & 0.00E+00 & 0.00E+00 \\
    17 & 2 & 0.1 & 1.172E-14 & 1.155E-13 & 2.423E-13 & 2.695E-17 & 1.126E-17 & 0.00E+00 & 0.00E+00 & 0.00E+00 \\
    17 & 2 & 0.2 & 3.532E-13 & 1.803E-12 & 4.542E-12 & 9.153E-15 & 2.943E-15 & 0.00E+00 & 0.00E+00 & 0.00E+00 \\
    17 & 2 & 0.4 & 1.937E-12 & 4.298E-12 & 1.243E-11 & 1.355E-13 & 6.473E-13 & 0.00E+00 & 0.00E+00 & 0.00E+00 \\
    17 & 2 & 0.8 & 4.742E-12 & 3.897E-12 & 1.234E-11 & 6.044E-13 & 9.377E-12 & 0.00E+00 & 0.00E+00 & 0.00E+00 \\
    17 & 2 & 1.6 & 7.735E-12 & 2.100E-12 & 7.103E-12 & 1.571E-12 & 1.991E-11 & 0.00E+00 & 0.00E+00 & 0.00E+00 \\
    17 & 3 & 0.1 & 1.598E-15 & 1.673E-14 & 4.488E-14 & 1.175E-18 & 2.620E-18 & 0.00E+00 & 0.00E+00 & 0.00E+00 \\
    17 & 3 & 0.2 & 4.130E-14 & 2.716E-13 & 9.077E-13 & 3.294E-16 & 6.841E-16 & 0.00E+00 & 0.00E+00 & 0.00E+00 \\
    17 & 3 & 0.4 & 1.600E-13 & 6.552E-13 & 2.614E-12 & 3.908E-15 & 3.143E-14 & 0.00E+00 & 0.00E+00 & 0.00E+00 \\
    17 & 3 & 0.8 & 2.288E-13 & 6.062E-13 & 2.771E-12 & 1.380E-14 & 1.321E-12 & 0.00E+00 & 0.00E+00 & 0.00E+00 \\
    17 & 3 & 1.6 & 1.888E-13 & 3.467E-13 & 1.755E-12 & 2.866E-14 & 2.700E-12 & 0.00E+00 & 0.00E+00 & 0.00E+00 \\
    17 & 4 & 0.1 & 9.519E-15 & 9.987E-14 & 1.652E-13 & 3.821E-18 & 7.047E-18 & 0.00E+00 & 0.00E+00 & 0.00E+00 \\
    17 & 4 & 0.2 & 3.010E-13 & 1.664E-12 & 3.200E-12 & 1.083E-15 & 1.847E-15 & 0.00E+00 & 0.00E+00 & 0.00E+00 \\
    17 & 4 & 0.4 & 1.674E-12 & 4.103E-12 & 8.932E-12 & 1.312E-14 & 7.561E-14 & 0.00E+00 & 0.00E+00 & 0.00E+00 \\
    17 & 4 & 0.8 & 4.101E-12 & 3.790E-12 & 8.889E-12 & 4.789E-14 & 4.173E-12 & 0.00E+00 & 0.00E+00 & 0.00E+00 \\
    17 & 4 & 1.6 & 6.688E-12 & 2.063E-12 & 4.957E-12 & 1.036E-13 & 9.098E-12 & 0.00E+00 & 0.00E+00 & 0.00E+00 \\
    17 & 5 & 0.1 & 1.847E-14 & 3.127E-14 & 3.764E-14 & 8.155E-18 & 3.337E-18 & 1.929E-09 & 1.814E-10 & 1.993E-10 \\
    17 & 5 & 0.2 & 5.173E-13 & 5.612E-13 & 7.604E-13 & 2.714E-15 & 8.774E-16 & 1.697E-09 & 1.460E-10 & 1.592E-10 \\
    17 & 5 & 0.4 & 2.080E-12 & 1.452E-12 & 2.218E-12 & 3.910E-14 & 1.955E-13 & 1.422E-09 & 1.146E-10 & 1.265E-10 \\
    17 & 5 & 0.8 & 3.026E-12 & 1.402E-12 & 2.387E-12 & 1.684E-13 & 1.998E-12 & 1.166E-09 & 9.029E-11 & 1.007E-10 \\
    17 & 5 & 1.6 & 2.522E-12 & 8.171E-13 & 1.525E-12 & 4.204E-13 & 4.891E-12 & 9.437E-10 & 7.147E-11 & 8.012E-11 \\
\hline
\end{tabular}
\begin{tablenotes}
\item[] Note: full-data table can be found via the link to the machine-readable version.
\item[$(a)$] Level index as given in Table~\ref{tab:levels}.
\item[$(b)$] Energy in unit of keV.
\item[$(c)$] Rate coefficient of direct collisional excitation from the ground without cascade.
\item[$(d)$] Rate coefficient of resonant excitation from the ground without cascade.
\item[$(e)$] Rate coefficient of resonant excitation from the ground including cascade.
\item[$(f)$] Rate coefficient of radiative recombination including cascade.
\item[$(g)$] Rate coefficient of dielectronic recombination including cascade.
\item[$(h)$] Rate coefficient of direct+resonant excitation from level 1 without cascade. 
\item[$(i)$] Rate coefficient of direct+resonant excitation from level 2 without cascade.
\item[$(j)$] Rate coefficient of direct+resonant excitation from level 3 without cascade.

\end{tablenotes}
\end{threeparttable}
\end{table*}

\subsection{Level population}

Here we evaluate the relative contribution of the various atomic processes to the line formation for a low-density plasma.
The level population is calculated using a built-in collisional-radiative program in SPEX, which solves the
occupation for each level directly with a large coefficient matrix. To separate different atomic processes, we run the program
several times, in each run we turn on only one of the five processes: direct collisional excitation, resonant 
excitation, dielectronic recombination, radiative recombination, and innershell ionization. The resonant excitation can be further 
divided into two components by the autoionizing doubly excited states. The rate coefficients of each process
to populate the upper levels of the target lines are recorded independently. All the data used in the line formation are calculated in 
this work, except for the radiative recombination rates which are based on the calculation in \citet{mao2016}.

It is well-known that many relevant levels, in particular those form the forbidden and intercombination lines, are significantly 
populated by radiatve cascades from higher states \citep{atomic2017}. In Fig.~\ref{fig:17apie}, we show the source compositions 
of the two \ion{Fe}{XVII} lines at $\sim$17~{\AA}. The cascade is clearly
the most important component, while the direct contribution is $\sim 20$\% of the total rates. Most of the cascades go through the 
$3s-3p$, $3s-3s$ ($^1$P$_{1}$ $-$ $^3$P$_{0}$), and $2s-2p$ transitions. It is therefore important to include the cascade component 
for each of the atomic processes. As shown in Fig.~\ref{fig:17atot}, the cascade-included rate coefficients of each process, for the four 2$p^5$3$s$ levels
of \ion{Fe}{XVII}, are calculated as a function of equilibrium temperature. It can be seen that the direct collisional excitation from the ground state
is the dominant process in $0.2-1.0$ keV, while the indirect excitation contributes $\sim 30$\%
of the $^3$P$_{2}$ population, and $\sim 10$\% of the other three states at 0.8~keV. The direct excitation populates these 
states mainly through cascades from levels at higher energies. The fractional contribution of indirect excitation increases
to $\sim 40-50$\% at 0.2~keV, as the resonant channels become relatively more efficient at a lower energy.

The results of the line formation calculation are recorded in Tables~\ref{tab:levels} and \ref{tab:rate}. The 
levels involved in the new calculation are listed in Table~\ref{tab:levels}. The notation is given in $LS$-coupling theme. 
Table~\ref{tab:rate} lists the temperature-dependent level-resolved rate coefficients for direct collisional excitation, resonant excitation,
radiative recombination, and dielectronic recombination. For the excitation, we include the rate coefficients from the ground state, and
those from three low-lying excited states. The printed version is truncated; tables with full data can be found as a machine-readable file in the electronic version.

\subsection{Spectra of the Fe-L complex}
\label{sec:felspectra}
\begin{figure*}[!htbp]
\centering
\resizebox{0.85\hsize}{!}{\includegraphics[angle=0]{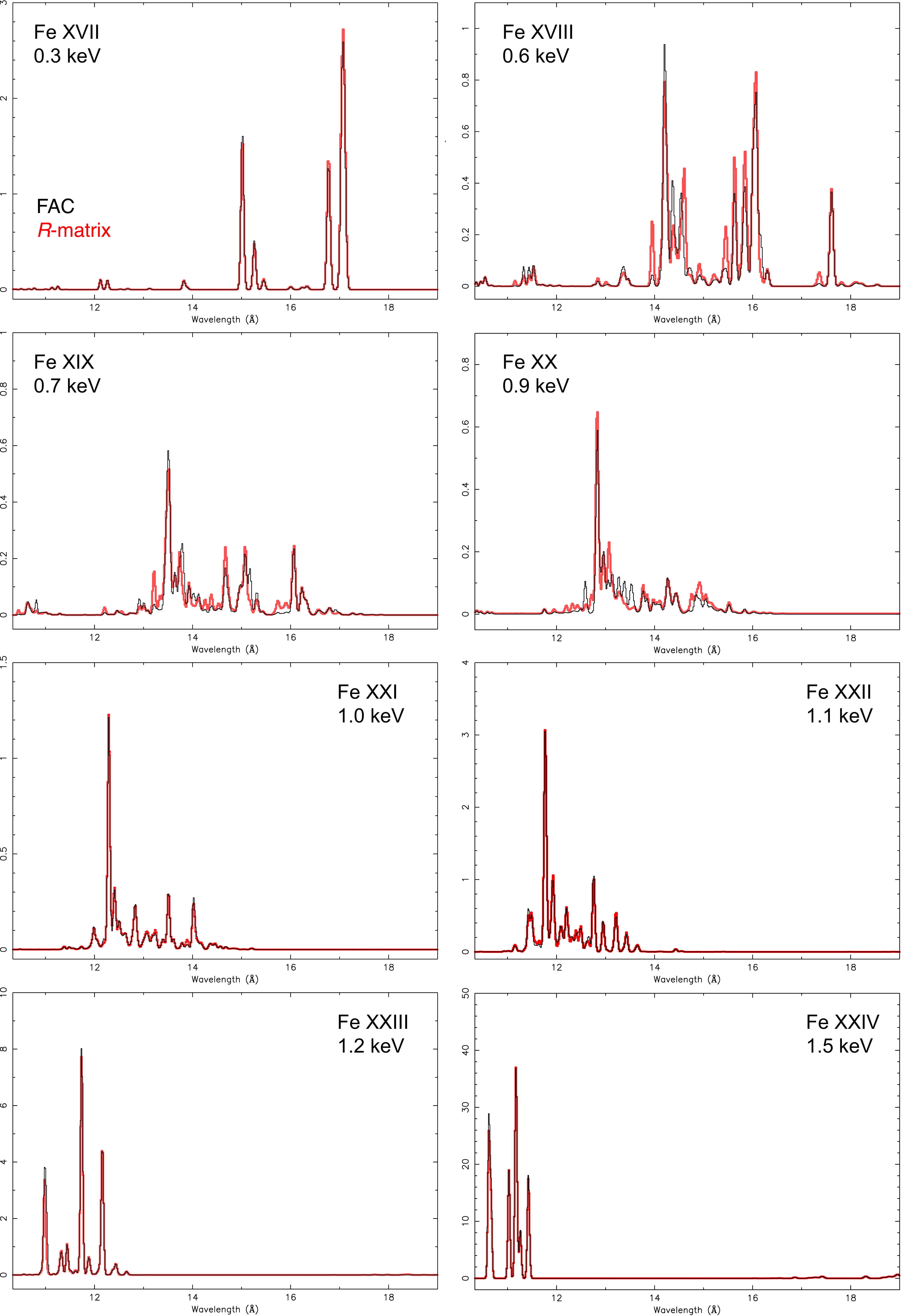}}
\caption{Model spectra of \ion{Fe}{XVII} to \ion{Fe}{XXIV} in the Fe-L band, obtained from the new FAC (black) and $R$-matrix (red) calculations.
The spectra are smoothed by a Gaussian with $\sigma \approx$ 2~eV, similar to the resolution of {\it Athena}. Temperature of each spectrum is set
to the value of peak ion concentration in equilibrium.}
\label{fig:facspec}
\end{figure*}

\begin{figure*}[!htbp]
\centering
\resizebox{0.85\hsize}{!}{\includegraphics[angle=0]{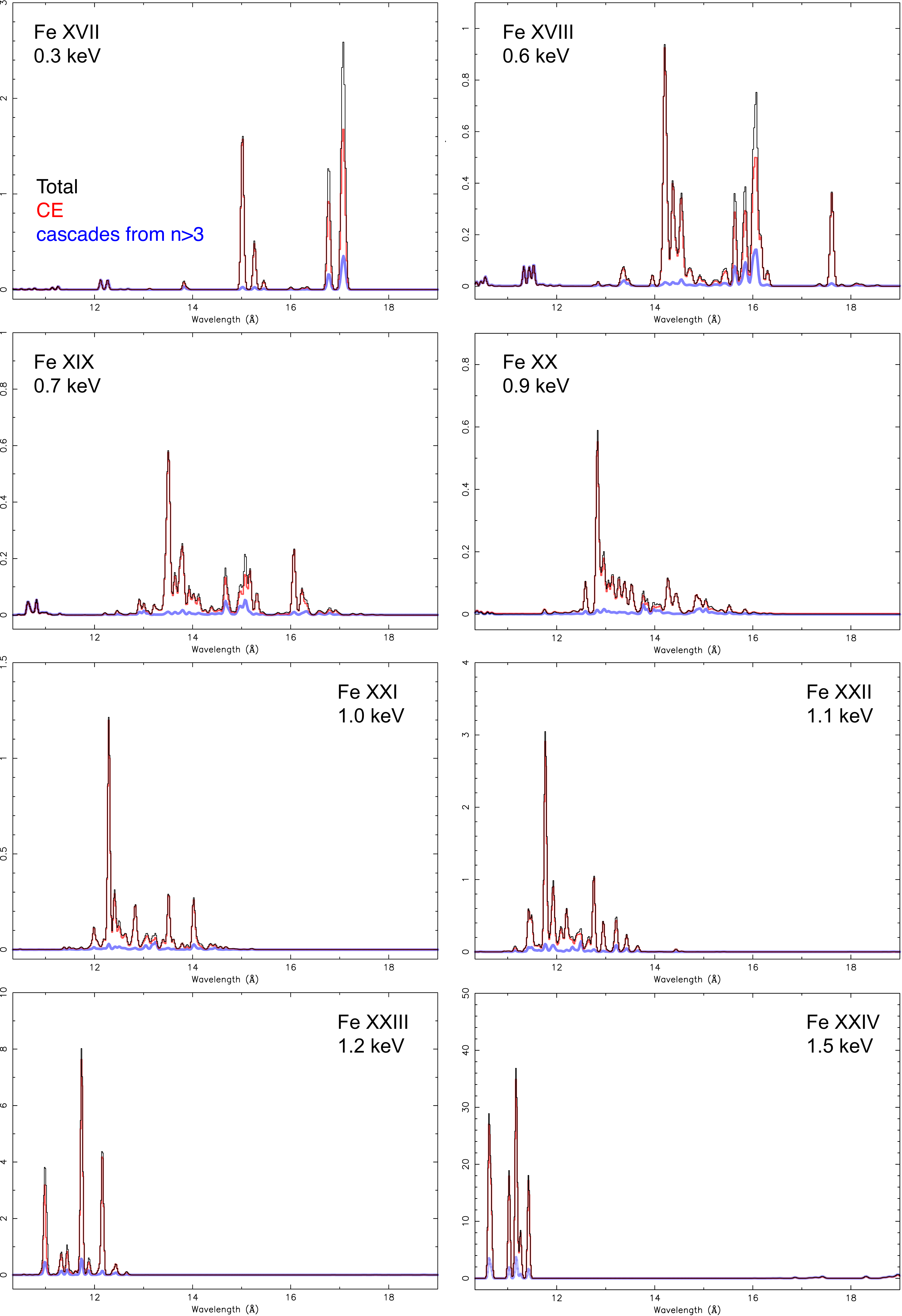}}
\caption{Model spectra of \ion{Fe}{XVII} to \ion{Fe}{XXIV} in the Fe-L band with the FAC calculation, highlighting the contribution from
direct collisional excitation including the cascades (red), and the contribution from the highly excited levels with $n > 3$ (blue).   }
\label{fig:facspec_cas}
\end{figure*}

The model spectrum for each Fe ion obtained from the current calculations is shown in Fig.~\ref{fig:facspec}. They are
compared with the models based on recent $R$-matrix collision calculations: \ion{Fe}{XVII} from \citet{liang2010}, \ion{Fe}{XVIII}
from \citet{witt2006}, \ion{Fe}{XIX} from \citet{butler2008}, \ion{Fe}{XX} from \citet{witt2007}, \ion{Fe}{XXI} from 
\citet{badnell2001}, \ion{Fe}{XXII} from \citet{liang2012}, \ion{Fe}{XXIII} from \citet{fern2014}, and \ion{Fe}{XXIV}
from \citet{liang2011}. The spectra are smoothed
to the resolution of the micro-calorimeter onboard {\it Athena} \citep{nandra2013}. The two sets of spectra are calculated using the same rate equation for solving the level population, and 
the input atomic data are the same except for the collisional excitation. Therefore, the differences can be interpreted as
the representative atomic uncertainties due to the theoretical modeling of the collision processes.

As shown in Fig.~\ref{fig:facspec}, the discrepancies between two codes, at temperatures of peak ion concentration in equilibrium, 
are mostly within 20\% on the main Fe-L transitions. The $R$-matrix
results give slightly higher emissivities for the \ion{Fe}{XVII} line at 17~{\AA} and the \ion{Fe}{XX} line at 12.8~{\AA}, while
the FAC calculation produces a higher \ion{Fe}{XVIII} transition at 14.2~{\AA} and a higher \ion{Fe}{XIX} line at 13.5~{\AA}. The
differences become significantly larger for the weaker transitions of \ion{Fe}{XVIII}, \ion{Fe}{XIX}, and \ion{Fe}{XX}. Similar results
can be found in \citet{fern2017}. As for \ion{Fe}{XXI} to \ion{Fe}{XXIV}, 
the two calculations agree within a few percent for all the main lines, as well as for most of the weaker ones. This comparison 
would help us to identify and prioritize the areas where laboratory measurements are needed to distinguish the theoretical models.

%Here we investigate further into one typical discrepancy: the FAC line intensity 
%for the \ion{Fe}{XVIII} line near 14.4~{\AA} is clearly higher than the $R$-matrix value, while the neighbour line near 14.6~{\AA}
%shows the opposite. This is probably because the FAC excitation rate to the level 2$s^2$2$p^4$3$d$ $^{2}$D$_{5/2}$ (14.37~{\AA}) is 
%twice as the $R$-matrix value, while the FAC rate to 2$s^2$2$p^4$3$d$ $^2$P$_{3/2}$ (14.61~{\AA}) is more than five times lower than
%the $R$-matrix one. It should be noted that the FAC line intensity near 14.6~{\AA} mostly comes from the transition of another level, 
%2$s^2$2$p^4$3$d$ $^2$F$_{5/2}$ (14.53~{\AA}), for which the FAC rate is nearly twice as the $R$-matrix value. The $^2$F$_{5/2}$ line will be
%blended with the $^2$P$_{3/2}$ line even for {\it Athena}. The changes in the blended line fluxes mimic a ``shift'' of the line peak,
%which is seen in the $R$-matrix-FAC comparison for \ion{Fe}{XVIII} near 14.6~{\AA}. Such ``shifts'' are also found for some 
%of the \ion{Fe}{XIX} and \ion{Fe}{XX} satellite lines. 

%It should be
%noted that, since many weak lines are heavily blended even for {\it Athena}, changes in the blended line fluxes might mimic small ``shifts'' of the 
%line peaks, although the wavelengths and atomic structures in the two calculations are identical. Such ``shifts'' are visible for 
%some of the \ion{Fe}{XVIII}, \ion{Fe}{XIX}, and \ion{Fe}{XX} lines. 

Figure~\ref{fig:facspec_cas} illustrates the contributions from different line-formation processes to the model spectrum obtained with the FAC calculation. 
This is achieved by a partial line formation calculation, including only a subset of atomic data for particular processes. The direct
collisional excitation with cascade is found dominant, at the temperature of peak ion concentration, for most lines in the Fe-L 
band. This confirms the results shown in Fig.~\ref{fig:17atot}. The cascade from highly excited levels ($n \geq 4$) has a moderate 
contribution. It is especially relevant for several lines,
e.g., the \ion{Fe}{XVII} lines at 16.80~{\AA}, 17.05~{\AA} and 17.09~{\AA}, the \ion{Fe}{XVIII} lines at 15.63~{\AA}, 15.83~{\AA}, 
and 16.07~{\AA}, the \ion{Fe}{XIX} lines at 14.67~{\AA} and 15.08~{\AA}, the \ion{Fe}{XX} line at 13.77~{\AA}, the \ion{Fe}{XXI} line
at 13.25~{\AA}, the \ion{Fe}{XXII} line at 12.50~{\AA}, the \ion{Fe}{XXIII} lines at 11.02~{\AA} and 11.74~{\AA}, and the \ion{Fe}{XXIV} 
lines at 10.62~{\AA}, 11.03~{\AA}, 11.17~{\AA}, and 11.43~{\AA}.

\subsection{Comparing with G03}
\label{sec:vg03}
\begin{figure*}[!htbp]
\centering
\resizebox{0.9\hsize}{!}{\includegraphics[angle=0]{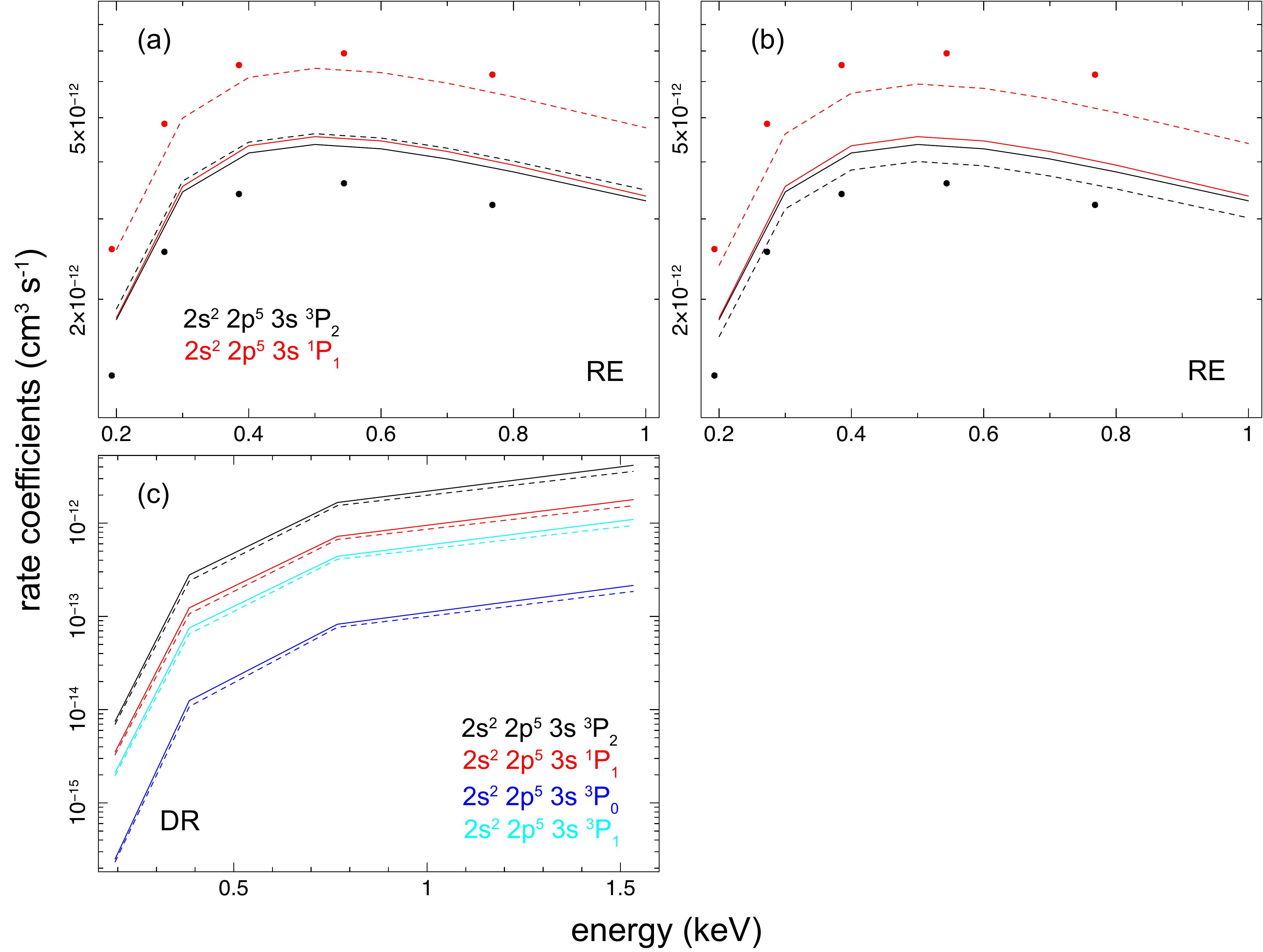}}
\caption{ (a) \ion{Fe}{XVII} resonant excitation rate coefficients for two low-lying levels, calculated with FAC version 1.1.4 (solid) and version 1.0 (dashed). 
The G03 results are shown in data points.  (b) \ion{Fe}{XVII} resonant excitation rates of the two levels, obtained with a full calculation 
with FAC version 1.1.4 (solid) and with a $l$-limited calculation with FAC version 1.0 (dashed). The latter can be compared directly with the G03
data points.
(c) \ion{Fe}{XVII} dielectronic recombination rates for four low-lying levels obtained with the ionization concentration data of \citet{u17} (solid) and 
those from G03 (dashed). }
\label{fig:vsg03}
\end{figure*}

The distorted wave calculation of G03 with the FAC code provided the rate coefficients of direct excitation, resonant excitation, dielectronic recombination,
radiative recombination, and innershell processes that populate the $n=2$ and $n=3$ states, for all the related L-shell species. Fits using the G03 data
to the {\it XMM-Newton} and {\it Chandra} grating spectra of Capella yielded a reasonable agreement \citep{gu2006}. To justify the updates of our work from G03, here we 
present a systematic comparison of the two papers.

\begin{enumerate}

\item[1] G03 calculated the collisional excitation only from the ground state. As shown in Appendix A and Table~\ref{tab:rate}, we consider both
the ground state and the low-lying excited states, as the latter is necessary for modeling intermediate-/high-density plasma.  

\smallskip

\item[2] Our calculation is done using the latest version of the FAC code, while G03 was based on an early version. In Fig.~\ref{fig:vsg03} (a), \ion{Fe}{XVII} resonant
excitation rates for two low-lying levels using the latest code (version 1.1.4) are compared with those calculated with the code version 1.0. The latest version 
gives lower resonant rates, by $\sim 5$\% for the 3$s$ $^3$P$_{2}$ level and $\sim 30$\% for the 3$s$ $^1$P$_{1}$ level, than the early version. 

\smallskip

\item[3]  As already noted in \S\ref{sec:intro}, G03 published the rate coefficients for a complete set of levels with $n=2$ and 3 in the paper. This contains the key transitions in the Fe-L 
complex, however, as shown in \citet{brickhouse2000}, the quantum number $n$ is still too low to sufficiently model the high-resolution spectra from bright X-ray coronal sources. 
The high-$n$ contributions are crucial for such sources.
To allow the test with real observational data (\S\ref{sec:apply}), in this work we calculate all the processes populating the states up to $n=5$.

\smallskip

\item[4]  The configurations of the doubly excited states (states $d$ in \S~\ref{sec:method}) are slightly different in two calculations. G03 limited their configurations up to $l' \leq 7$ 
for 3$lnl'$, and $l' \leq 4$ for 4$lnl'$, while we include all possible configurations for each $n$. Naively, the resonant excitation rate coefficients will increase
by the additional doubly excited levels. For the two \ion{Fe}{XVII} test levels shown in Fig.~\ref{fig:vsg03} (b), the resonant rates using the $l-$limited
calculations are indeed lower, by $\sim 10-15$\%, than those obtained in the complete calculation. As the $l-$limited 
rate coefficients shown in Fig.~\ref{fig:vsg03} (b) are obtained with FAC version 1.0, they could be compared directly with the G03 results.
It appears that the two sets of rates still differ by $5-20$\%, suggesting that there are other sources of discrepancy in the calculation.

\smallskip

\item[5]  According to Eq.\ref{eq:dr}, the different ionization balance used in the two calculations might introduce discrepancies to the dielectronic recombination rates. To quantify the
effect, we apply the ionization balance from G03 and calculate the rates again for the \ion{Fe}{XVII} test levels. As shown in Fig.~\ref{fig:vsg03} (c), the rates with
G03 ionization balance are lower by $\sim 8$\% than the rates with the balance from \citet{u17}. This is because the \ion{Fe}{XVIII} to \ion{Fe}{XVII} ratios
in the new ionization balance standard are slightly higher than those in G03. 

\smallskip

\item[6]  G03 calculated the level populations in a hierarchical way. First, a large number of levels were grouped into super levels. The overall population of each super level 
was calculated. It was then partitioned into each level within the group. In our work, the populations of all levels are solved at once using a large coefficient matrix. 
As reported in \citet{lucy2001}, the super level method applying to a system with $\sim 1000$ levels can reach an accuracy of 0.1 with 6 iterations, and 0.01 with 20 iterations. 
Meanwhile, \citet{poirier2007} showed that the super level method, as adopted in G03, might become less accurate when the rms deviation of transition rates inside one super level increases. 

\end{enumerate}

To summarize, we prove that the new theoretical calculation has become both more accurate and more complete than the pioneering G03 calculation.

\section{Application to high-resolution X-ray grating data}
\label{sec:apply}

Here we test the new Fe-L calculations on high spectral resolution X-ray data of celestial objects. 
The targets are selected to be the intracluster medium (ICM) of bright nearby elliptical galaxies/galaxy clusters.
They can be regarded as an ideal plasma in collisional ionization equilibrium thermalized to a balance temperature
of $\sim 0.5-1.5$ keV, and enriched to near-Solar abundances \citep{mernier2017, hitomi2017}. Although the X-ray fluxes
of the ICM sources are substantially lower than those of the coronae of nearby stars (e.g., Capella), they are in general 
astrophysically simpler, as the temperature components mixed into the ICM emission model are apparently fewer
than those of the stellar coronae.

%Two groups of targets 
%are selected, bright nearby ellliptical galaxies/galaxy clusters, and the bright non-degenerate stellar corona of the G1 + G8 binary 
%Capella. The primary X-ray source of the first group is the intracluster medium (ICM), which
%can be regarded as an ideal plasma in collisional ionization equilibrium thermalized to a balance temperature
%of $\sim 0.5-1.5$ keV, and enriched to near-Solar abundances \citep{mernier2017, hitomi2017}. The quiescent coronal plasma of Capella
%is also known to be in a collisional equilibrium state, heated to a continuous temperature distribution over the range of $10^{5} - 10^{7}$ K
%\citep{dupree1993, brickhouse2000, p2001, behar2001, desai2005, gu2006}. Although the X-ray fluxes of the ICM sources are substantially lower than
%that of Capella, they are in general astrophysically simpler, as the temperature components mixed into the ICM emission model are apparently fewer
%than those of Capella. The atomic tests with these two types of objects are well complementary to each other. 

The main purpose of the testing is to reveal the possible biases and systematic uncertainties on the key source parameters due to the change of the underlying atomic database.
Three databases with different Fe-L calculations are established: the default data in SPEX version 3.04 are used as the first model (hereafter model 0), which include 
distorted wave calculations of the direct collisional excitation, and a limited set of dielectronic recombination rates for the Fe-L species (\S\ref{sec:method_dr}).  
The second model, hereafter model 1, includes a complete set of the new calculations done in this work. We also construct the third model (hereafter model 2) by implementing the recent 
$R$-matrix calculations for the collisional excitation (see the list in \S\ref{sec:felspectra}). The atomic structure, radiative recombination, dielectronic recombination, and the
innershell data of model 1 and model 2 are the same.

In principle, model 0 should be the least accurate among the three due to the incomplete resonance channels, though it is currently widely used in X-ray astronomy 
\citep{hitomi2017, atomic2017, anna2017, mernier2016, mer2016, mernier2017, mao2018}. Model 1 and model 2 
should have similar quality, though they are still different in many places (Fig.~\ref{fig:facspec}). Comparing the astronomical measurements using model 0 with the other two will indicate the 
possible biases in the
previous results reported in literature. The difference between the model 1 and model 2 results can be used as a rough estimate of the representative systematic uncertainties from atomic databases.

Note that we do not intend to verify the new calculation with the observed data. In fact, none of the current observational data allows a full validation of 
the atomic database. The astrophysical effects, such as the differential emission measure distribution and the resonant scattering, along with the 
common assumptions made in analysis (e.g., uniform abundances for all temperature components), might hamper an accurate 
benchmark of the atomic database. The data from controlled laboratory experiments \citep{brown2006} are clearly more suited for such a purpose.

%Here we test the new Fe-L calculations on the high-spectral-resolution X-ray observations of bright nearby elliptical galaxies and
%galaxy clusters. The primary X-ray source of these objects is the intracluster medium (ICM), which
%can be regarded as an ideal plasma in collisional ionization equilibrium thermalized to a balance temperature
%of $\sim 0.5-1.5$ keV, and enriched to near-Solar abundances \citep{mernier2017, hitomi2017}. Although the
%grating data from the quiescent corona of stars (e.g., Capella) generally provide a better spectral quality, 
%the stellar plasma often exhibits complicated astrophysical effects, such as the differential emission measure 
%distributions \citep{gu2006}, making the data less straightforward for testing atomic modeling.  
%The ICM in ellipticals and clusters is in general astrophysically simpler. Comparing to the coronal temperature, 
%the mean ICM temperatures spread a wider range in different objects, providing a better test of the atomic calculation
%at different ionization states. Therefore the ICM is chosen for calibrating the modeling of the Fe-L complex.

\begin{figure*}[!htbp]
\centering
\resizebox{0.9\hsize}{!}{\includegraphics[angle=0]{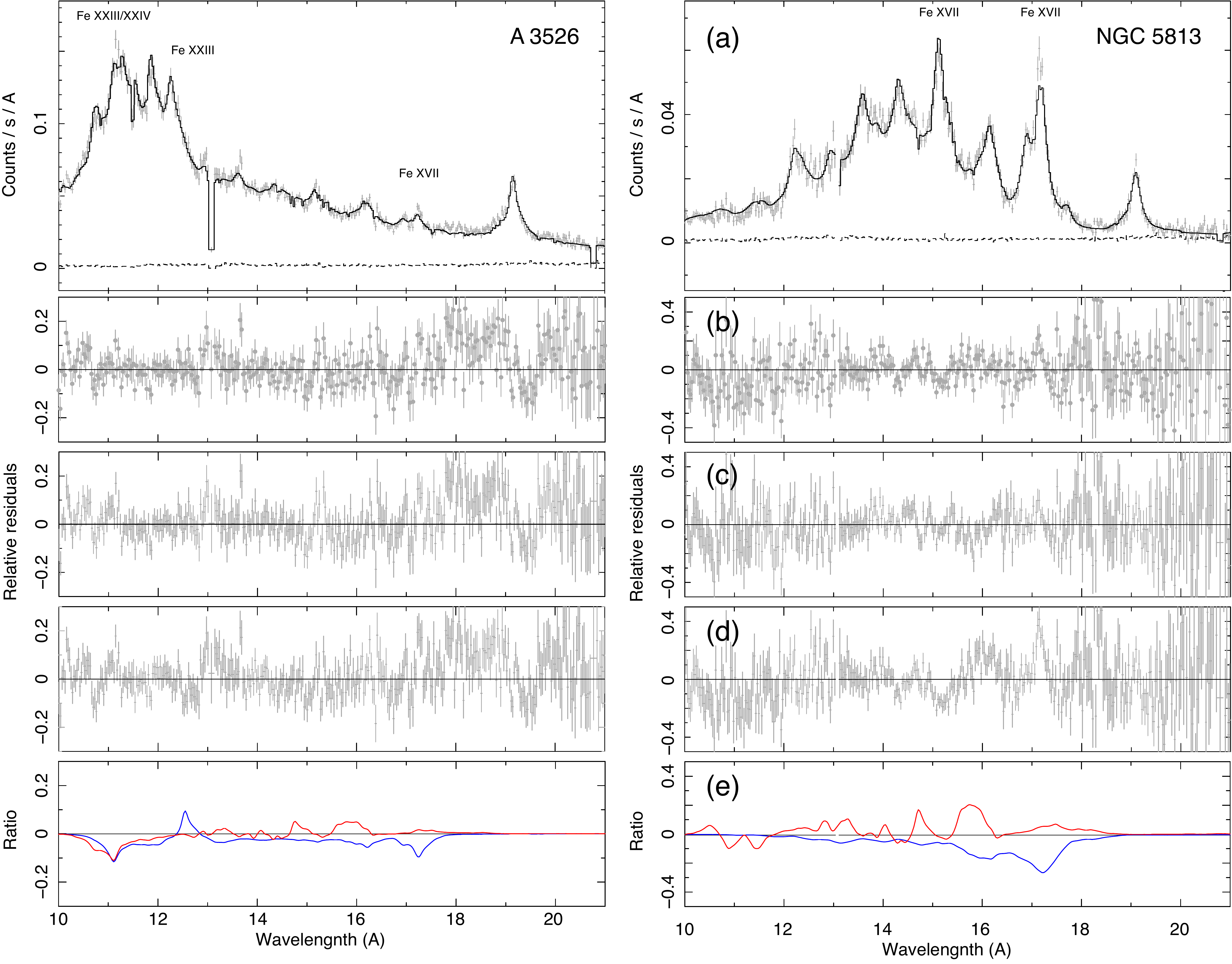}}
\caption{RGS spectra of the central 3.4-arcmin regions of Abell~3526 (left) and NGC~5813 (right) in the $10-21$~{\AA} band fitted with different models. Panels (a) show the fits by the 
two-temperature {\it cie} with model 1, the residuals are shown in panels (b). Panels (c) and (d) show the residuals of the fits with models 2 and 0, respectively. Panels (e) show the 
ratios among the three model spectra. The model 0 to model 1 ratios are plotted in blue, and the model 2 to model 1 ratios are plotted in red. 
It could be seen that the line emissivities of model 1 are higher than those of model 0, but slightly lower than those of model 2, in the $15-17$~{\AA} band.  }
\label{fig:5813}
\end{figure*}

\begin{figure*}[!htbp]
\floatbox[{\capbeside\thisfloatsetup{capbesideposition={right,center},capbesidewidth=7cm}}]{figure}[\FBwidth]
{\caption{Relative difference of the best-fit C-statistic values with model 1 and model 0 as a function of temperature. 
The temperatures are obtained using the single-temperature fits with model 1.}\label{fig:cstat}}
{\includegraphics[width=11cm]{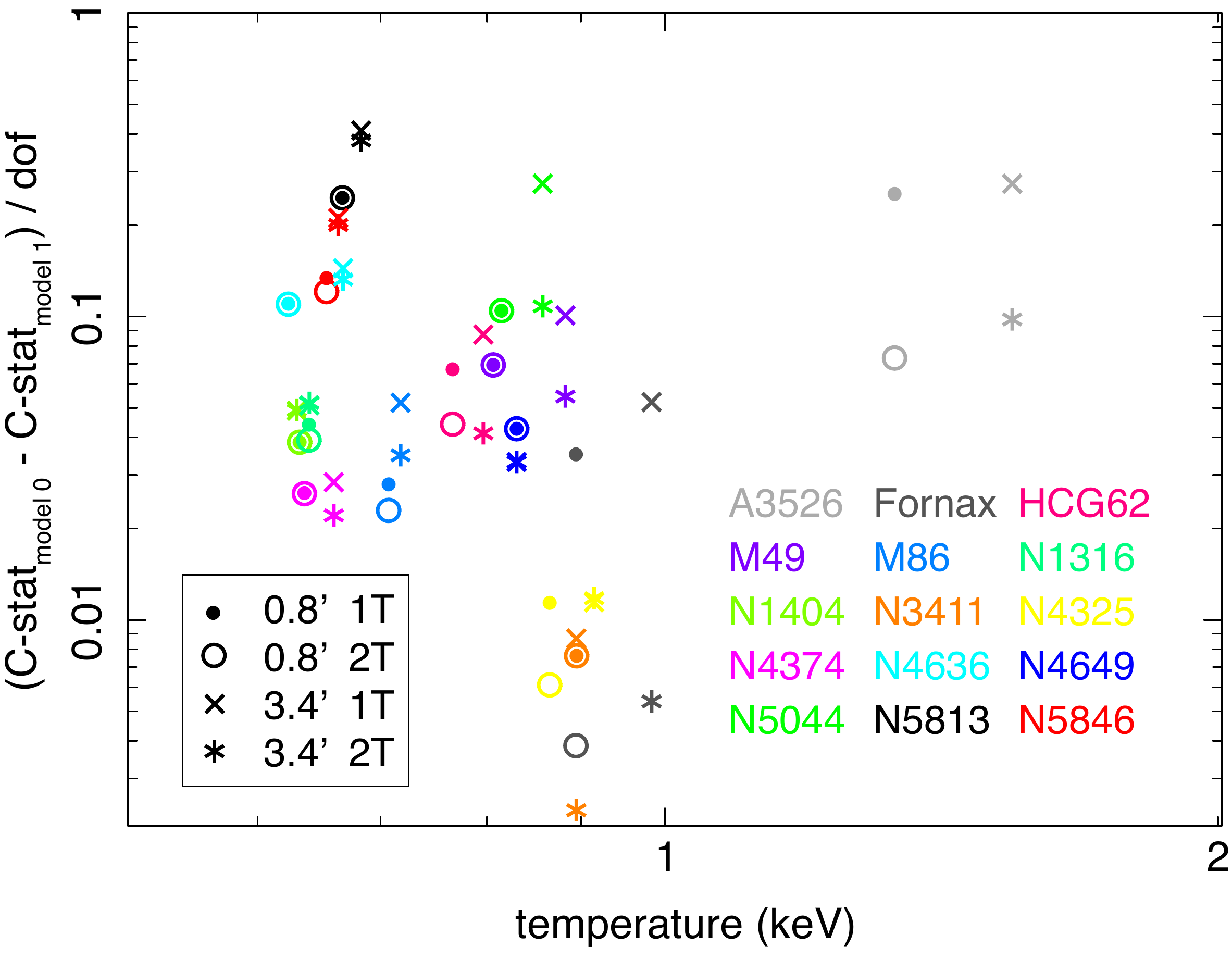}}
\end{figure*}

\begin{figure*}[!htbp]
\centering
\resizebox{0.9\hsize}{!}{\includegraphics[angle=0]{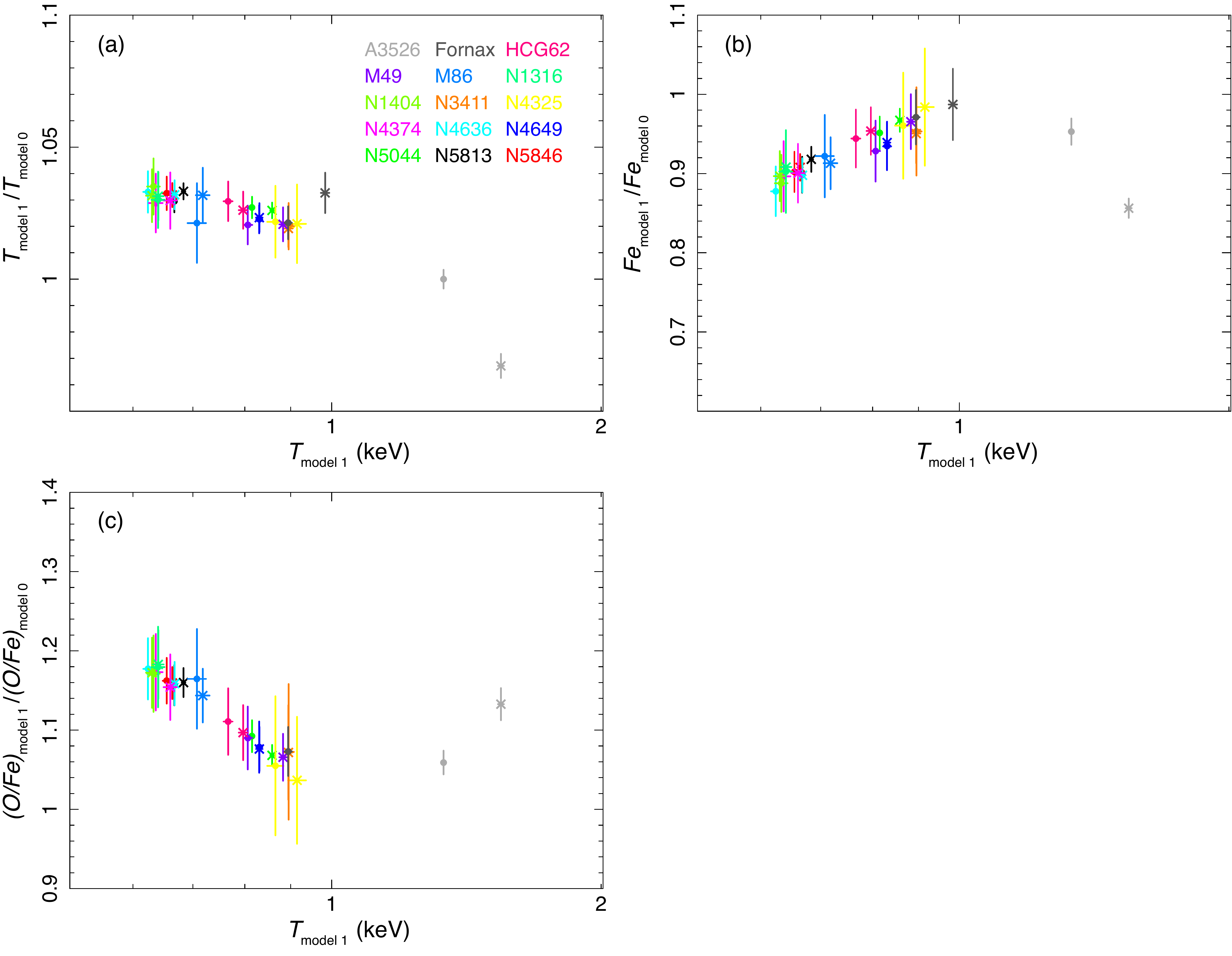}}
\caption{Ratios of the best-fit (a) temperatures, (b) Fe abundances, and (c) O/Fe with model 1 and model 0. 
All values are obtained in the single-temperature fits. Data with crosses and dots are ratios determined in the 3.4~arcmin
and 0.8~arcmin regions, respectively.}
\label{fig:1tcom}
\end{figure*}

\begin{figure*}[!htbp]
\centering
\resizebox{0.9\hsize}{!}{\includegraphics[angle=0]{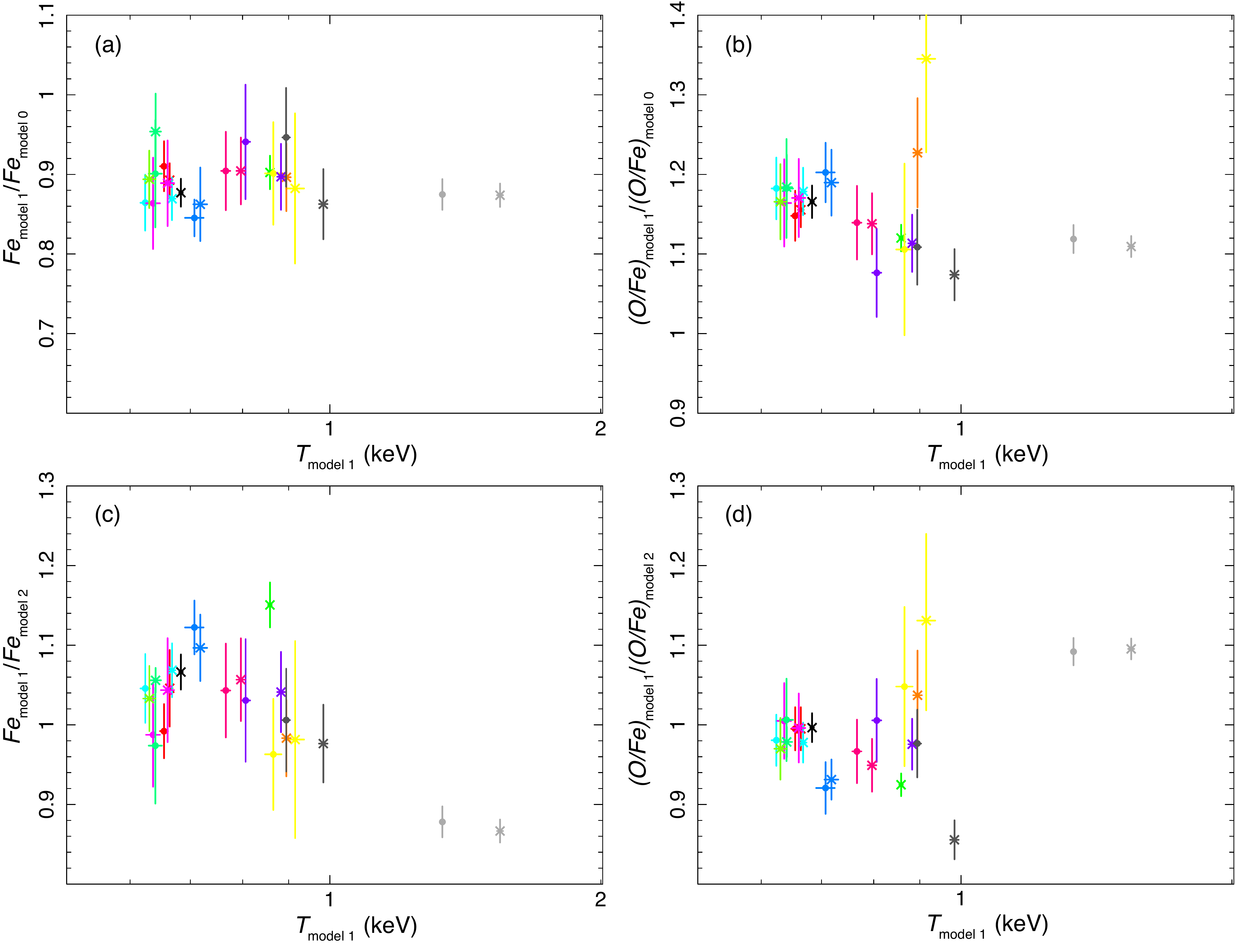}}
\caption{Ratios of the best-fit (a) Fe abundances and (b) O/Fe with model 1 and model 0, obtained in the two-temperature
fits. Panels (c) and (d) are the ratios of Fe abundances and O/Fe with model 1 and model 2 in the two-temperature fits.
The color theme are the same as Fig.~\ref{fig:1tcom}.}
\label{fig:2tcom}
\end{figure*}

\subsection{{\it XMM-Newton} grating data }
\label{sec:analysis}

Among the current X-ray observatories, the Reflection Grating Spectrometer (RGS, \citealt{jw2001}) onboard {\it XMM-Newton} 
has the unique power to resolve the Fe-L emission from the ICM into individual lines. The RGS spectra have been
used for measuring the chemical abundances of the ellipticals and galaxy clusters \citep{dp2017}, determining the turbulence velocity \citep{dp2012, pinto2015, anna2017},
and even probing weak non-thermal charge exchange emission lines \citep{gu2018, gu2018b}. These measurements are all sensitive to the underlying atomic modeling.
Here we apply our new calculation to a sample of RGS data of nearby elliptial galaxies and clusters.

All the testing objects are selected from the CHEmical Evolution RGS Sample (CHEERS), which is made up of 44 
representative nearby X-ray bright galaxies and clusters \citep{dp2017}. In this work, we focus on objects showing strong 
\ion{Fe}{XVII} lines in the spectra. The RGS study of \citet{pinto2016} had the same research focus, leading them 
to select a subsample of 24 objects. For the 24 objects, we further remove those with data of poor spectral quality. 
Objects with very diffuse morphology, such as M87, are not included in the final sample, as their spectra suffer 
too much from the instrumental broadening. The final sample consists of 15 objects. The properties of the selected targets are listed in
Table~\ref{tab:sample}.

\begin{table*}[!htbp]
\centering
\caption{\textit{XMM-Newton} RGS data}
\label{tab:sample}
\begin{threeparttable}
\begin{tabular}{lccccccccccc}
\hline
cluster     & Observation ID                     & Total clean time (ks)   & $kT^{(a)}$ (keV) & $z^{(a)}$ & $N_{\rm H}^{(b)}$ (10$^{24}$ m$^{-2}$)  \\
\hline 
Abell~3526      & 0046340101 0406200101              & 139.1                   & 3.7              & 0.0103     & 8.43 \\
Fornax      & 0012830101 0400620101              & 121.9                   & 1.2              & 0.0046     & 2.56 \\
HCG~62      & 0112270701 0504780501 0504780601   & 118.3                   & 1.1              & 0.0140     & 4.81 \\
M49         & 0200130101                         & 58.5                    & 1.0              & 0.0044     & 2.63 \\
M86         & 0108260201                         & 41.9                    & 0.7              & -0.0009    & 3.98 \\
NGC~1316    & 0302780101 0502070201              & 121.5                   & 0.6              & 0.0059     & 1.90 \\
NGC~1404    & 0304940101                         & 27.7                    & 0.6              & 0.0065     & 1.57 \\
NGC~3411    & 0146510301                         & 15.6                    & 0.8              & 0.0152     & 4.25 \\
NGC~4325    & 0108860101                         & 14.2                    & 1.0              & 0.0259     & 3.54 \\
NGC~4374    & 0673310101                         & 69.3                    & 0.6              & 0.0034     & 3.38 \\
NGC~4636    & 0111190101/0201/0501/0701          & 80.8                    & 0.8              & 0.0037     & 1.40 \\
NGC~4649    & 0021540201 0502160101              & 86.4                    & 0.8              & 0.0037     & 2.23 \\
NGC~5044    & 0037950101 0554680101              & 110.5                   & 1.1              & 0.0090     & 7.24 \\
NGC~5813    & 0302460101 0554680201/0301/0401    & 129.7                   & 0.5              & 0.0064     & 3.87 \\
NGC~5846    & 0021540101/0501 0723800101/0201    & 131.0                   & 0.8              & 0.0061     & 4.26 \\

\hline
\end{tabular}
\begin{tablenotes}
\item[$(a)$] Temperatures and redshifts are taken from \citet{dp2017}.
\item[$(b)$] Hydrogen column density are taken from \citet{mernier2016}.
\end{tablenotes}
\end{threeparttable}
\end{table*}

We process the {\it XMM-Newton} RGS and MOS data, following the method described in \citet{gu2018}.
The MOS data are used for screening soft proton flares and for deriving the spatial extent of
the source along the dispersion direction of the RGS detector.

The Science Analysis System (SAS) v16.1.0 and the latest calibration files (March 2018) are used for data reduction. 
The time interval contaminated by soft protons are identified using the lightcurves of the RGS CCD9 and 
the MOS data. The flaring periods are filtered out by a 2$\sigma$ clipping. For each object, two source spectra 
are extracted from a $\sim 3.4$-arcmin-wide belt and a $\sim 0.8$-arcmin-wide belt centered on the emission peak.
The modeled background spectra are used in the spectral analysis. 

Since RGS is a spectrometer without a slit, the source spatial extent causes the spectral features to be broadened. 
To model the broadening, we extract the MOS1 image in detector coordinate in $7-30$~{\AA}, and calculate the surface 
brightness profile in the RGS dispersion direction. The brightness profiles are convolved with the model spectrum
using the SPEX model $lpro$.

\subsection{Spectral modeling}

We analyze the first order RGS1 and RGS2 spectra in the $7-30$~{\AA} band. The metal abundances are scaled to the proto-Solar
standard of \citet{lodders2009}, and the Galactic absorption column densities are taken from \citet{mernier2016}.
The new ionization balance calculation presented in \citet{u17} is applied. The best-fit parameters are obtained by minimizing the C-statistics.

The dominant thermal component of the targets is first modeled with a {\it cie} component in SPEX. This is sometimes
inadequate, as many of the targets show both hot and cool gas phases \citep{frank2013}. Therefore, we also fit the data with two
{\it cie} models of different temperatures. Free parameters of the thermal components are the emission measure, the temperature, the abundances
of N, O, Ne, Mg, Fe, and Ni, and the velocity of the micro turbulence. In the case of two-{\it cie}, the abundances and 
the turbulent velocities of the two gas phases are bound to each other. As shown in Fig.~\ref{fig:cstat}, the two temperature
fits are in general better than the single temperature one. For a few objects, such as NGC~1404 and NGC~3411, the C-statistics
differences between the two temperature fits and the single temperature fits are small, as the second thermal component appears to be
weak.

\subsection{Biases and systematic uncertainties in the abundance measurement}

%The {\it cie} model in SPEX is backed by the atomic database for ionization concentration, continuum emission,
%and various line-formation processes. The new data of the Fe-L emission (\S\ref{sec:result}) are incorporated in the 
%line formation calculation, while the database for concentration, continuum, and line emission of other species are kept the same
%as the present code. This forms a modified {\it cie} model.
%Independent fits are carried out for the original {\it cie} in SPEX v3.04 and the modified {\it cie} models.

We implement model 0, model 1, and model 2 to the SPEX software and create three different versions of {\it cie}. For each object, the RGS spectrum
is fit independently using the three different {\it cie} versions. 
Fig.~\ref{fig:5813} plots two representative spectra of Abell~3526 and NGC~5813 fit with the three models. The two spectra reveal different ionization states,
as the mean temperatures are 0.64~keV and 1.6~keV for NGC~5813 and Abell~3526, respectively. 
The two temperature structure is taken into account in the fits. It shows that model 1 and model 2 improve significantly relative to model 0
in the fits for NGC~5813. The \ion{Fe}{XVII} lines at 17{\AA} contribute significantly to the fit improvement, as they are clearly affected by the 
new resonant excitation and dielectronic recombination data (Fig.~\ref{fig:17atot}). The improvements on the fits of Abell~3526 are less apparent. The ratio plots
show that the maximum discrepancies of the three models are about 10\% on \ion{Fe}{XXIV}, \ion{Fe}{XXIII}, and \ion{Fe}{XVII} lines for Abell~3526,
and about 20\% on \ion{Fe}{XVII} lines for NGC~5813.

%The results for the entire sample are summarized in Table~\ref{tab:rgsresult}, the changes in the best-fit statistics are shown in Fig.~\ref{fig:cstat},
%and the differences on the best-fit temperature, Fe abundance, and O/Fe of the new model relative to the original one are
%plotted in Figs.~\ref{fig:1tcom} \& \ref{fig:2tcom}. 

Fig.~\ref{fig:cstat} demonstrates that model 1 always gives better fits than model 0. On average,
the C-statistic value of the 3.4-arcmin region is improved by 86 (single-temperature) and 64 (two-temperature) for mean degrees of freedom of 713.
For the core 0.8-arcmin region, the mean statistics are improved by about 71 (single-temperature) and 44 (two-temperature) for the same degrees of freedom. The 
resulting $\Delta$C shows a weak dependency on the best-fit temperatures: the fits of the objects with $kT \sim 0.9-1.0$ keV are
less affected by the atomic data update than those with lower or higher temperatures. Model 2 provides a similar improvement on the fit statistics:
the average C-statistic values reduce by 84 (single-temperature) and 61 (two-temperature) from the model 0 fits for the 3.4-arcmin region.
It is not possible to distinguish between model 1 and model 2 with the current fits.

Figs.~\ref{fig:1tcom} and \ref{fig:2tcom} show a mild bias in temperature and abundance measurement due to the atomic data update. For the single-temperature
modeling, the sample-average temperature increases by 2\%, and the Fe abundance decreases by 9\% by switching from model 0 to model 1.
The ratio of O to Fe abundances would increase by a larger mean value of 13\%, as the changes in temperature and Fe abundance would affect indirectly
the O abundance (even though the atomic data for oxygen ions remain the same). The O/Fe ratio is a key parameter for quantifying the relative enrichment contribution from different types of supernovae to the interstellar and intracluster medium \citep{dp2017}. The biases on the abundances are larger at $\leq 0.8$ keV and potentially also significant at $\geq 1.3$ keV,
while the best-fit values around $\sim 1$ keV obtained with the SPEX v3.04 code might require just a minor revision. 

As for the two-temperature astrophysical modeling, the average Fe abundance decreases by 12\% with model 1, and the mean O/Fe ratio
increases by 16\%, relative to that obtained with model 0. These differences are slightly larger than the single-temperature cases. As shown in 
Fig.~\ref{fig:2tcom}, the changes on the Fe abundances show very weak dependence on the temperature. The O/Fe ratios still vary with temperature: a higher bias of $\sim 20$\% 
is found at $\leq 0.8$ keV, while the bias at $\geq 1$ keV becomes slightly lower. Considering that the two-temperature is naturally a better recipe for the cool-core objects 
than the single-temperature one \citep{gu2012}, the biases found in the two-temperature fits should be a better approximation to the reality.

As shown in Fig.~\ref{fig:2tcom}, the Fe abundances measured with model 2 appear to deviate from the model 1 results. 
The observed discrepancies seem to change as a function of temperature: the mean Fe abundance with model 1 is higher by $\sim 10$\% at 0.7~keV, but it
becomes lower by $10$\% at 1.5~keV, than the mean model 2 abundance. Current RGS data cannot decisively distinguish 
between model 1 and model 2 by the fit statistics, therefore, the 10\% abundance differences
can be treated as systematic uncertainties.
Furthermore, taking into account the model 1 versus model 0 ratios (Fig.~\ref{fig:2tcom}), the Fe abundances with
model 2 are lower than the model 0 values by $\sim 20$\% at $\sim 0.7$~keV, while the difference becomes smaller as the temperature increases (or decreases), and largely diminishes at 1.5~keV. The 
mean O/Fe ratio measured with model 2 is 23\% higher than the model 0 value below 1~keV, and the two values converge at 1.5~keV. 

The bias in measuring O/Fe ratio could affect the fraction of type~Ia supernovae contributing to the ICM enrichment (see the reviews of \citealt{boh2010} and \citealt{mernier2018}). 
As shown in \citet{simi2009}, the increase of 23\% in the O/Fe ratio might lead to a lower type~Ia fraction by $\sim 5-15$\%, depending on
the supernova explosion mechanism. The improved abundance ratio measurement can, in principle, also better distinguish among the type~II supernovae 
models with different level of pre-enrichment of the progenitors and with different initial-mass functions \citep{mer2016}.

This experiment provides a general idea of the spectroscopic sensitivity on the new Fe-L atomic calculations, for RGS spectra of a limited sample of elliptical 
galaxies and cool clusters with temperatures of $0.6 - 1.5$ keV. To summarize, for the cool objects ($< 1$ keV), the Fe abundances measured with the new calculations (model 1 and model 2) 
are consistently lower, by 10\% $-$ 20\%, than those derived from the standard plasma code (model 0). The systematic uncertainties on the Fe abundances, determined by comparing the 
model 1 and model 2 fits, are up to $10$\% for the current observations. 

The test is far from complete, as the new calculations still need to be tested on further cooler ($< 0.6$ keV) 
or hotter ($> 1.5$ keV) objects in CIE, non-equilibrium ionization objects, as well as the objects affected by a strong photon field. On the other
hand, the current RGS spectra resolve mostly the main transitions, while the satellite lines, which are strongly affected by the new atomic database,
cannot be fully tested. We expect that the new high-resolution X-ray spectrometers on board the X-ray imaging and spectroscopy mission ({\it XRISM}, \citealt{tashiro2018})
and {\it Athena} will be able to provide a sufficient test to these weak lines.

\/*
\begin{figure*}[!htbp]
\floatbox[{\capbeside\thisfloatsetup{capbesideposition={right,center},capbesidewidth=7cm}}]{figure}[\FBwidth]
{\caption{Observed and calculated values of the \ion{Fe}{XVII} 17~{\AA} / 15~{\AA} line ratio as a function of temperature. 
The observed values are from the fits of the central 0.8-arcmin regions, the color theme are the same as Fig.~\ref{fig:1tcom}. 
For Abell~3526, the temperature is assumed to be the cool phase temperature in the two-temperature model. The lines indicate
the values predicted by the spectral codes: SPEX v3.04 or model 0 (dashed), model 1 (solid black), model 2 (solid blue) and AtomDB/APEC 
v3.0.9 (solid red). The black triangle shows the laboratory measurement result taken from \citet{bei2004}.}\label{fig:gausst}}
{\includegraphics[width=11cm]{gauss_t.eps}}
\end{figure*}

\subsection{\ion{Fe}{XVII} line ratios}

In addition to the fits with a global spectral model, astrophysical/laboratory spectroscopic analysis is 
sometimes done with a local fit. It models a segment of a highly-resolved spectrum approximately as a sum 
of a simple continuum plus a series of Gaussian lines, with physical parameters derived from positions,
widths, and fluxes of the lines. This method is often used in the search for resonant scattering in ellipticals and clusters. The
most studied lines are the \ion{Fe}{XVII} main transition lines at 15.01~{\AA} ($2s^{2}2p^{5}3d$ $^1$P$_{1}$
to the ground), 17.05~{\AA} ($2s^{2}2p^{5}3s$ $^1$P$_1$ to the ground), and 17.09~{\AA} ($2s^{2}2p^{5}3s$ $^3$P$_2$
to the ground), which are all well-detected lines in elliptical galaxies with RGS \citep{xu2002, werner2009, dp2012}.
The 15.01~{\AA} transition has the largest oscillator strength, while the 17.05~{\AA} and 17.09~{\AA} lines
are much weaker, making the line ratio ($I_{17.05} + I_{17.09}$) / $I_{15.01}$ an excellent indicator for the resonant 
scattering. As shown in Fig.~\ref{fig:17atot}, the 17.05~{\AA} and 17.09~{\AA} line intensities are strongly affected by
the resonant excitation and dielectronic recombination. It is hence important to see how do the new atomic data
revise the line ratio for the resonant scattering study, and how do they match with the observed data.

Here we carry out another fit to the RGS data, focusing on the \ion{Fe}{XVII} lines. The spectra from
the central 0.8-arcmin regions are used, as the line opacity should be largest in the innermost regions.  
The spectra are fit with a single-temperature model, except for Abell~3526 which clearly requires a second 
temperature component. The fit is run with model 1.
The 15.01~{\AA}, 17.05~{\AA}, and 17.09~{\AA} \ion{Fe}{XVII} lines are omitted in the $cie$ model, replaced
by three Gaussian components in the fits, all with fixed wavelengths. The lines are broadened instrumentally by the 
spatial profiles derived from the MOS1 data, and their intrinsic width are set free. The other \ion{Fe}{XVII}
lines are included in the $cie$ model.
The new fit would allow an accurate measurement of the line ratio, largely independent of the spectral code.
The observed line ratios as a function of the best-fit temperatures, as well as the theoretical values from model 0,
model 1, model 2, 
and the latest AtomDB/APEC code, are plotted in Fig.~\ref{fig:gausst}.

The new calculations (models 1 and 2) match within error with the tokamak laboratory measurement reported in \citet{bei2004},
which is believed to be free from resonant scattering. The standard SPEX v3.04 code (model 0) predicts a much lower 
line ratio than the new calculations, probably due to the insufficient atomic data for resonant excitation and dielectronic recombination
in model 0. The discrepancies on the line ratios between model 1 and model 2 are 15\% at 0.2~keV, and 8\% at 1~keV.

As shown in Fig.~\ref{fig:gausst}, the latest AtomDB/APEC and model 1 predict the same temperature dependence in the 17~{\AA} to 15~{\AA}
line ratio, while APEC is systematically higher than our calculation by a near-constant of $\sim 12$\%. The model 2 results agree well
with AtomDB/APEC at low temperatures, and become lower than the latter by $\sim 5$\% at $\sim 1$~keV. The AtomDB/APEC values are 
also consistent with the tokamak measurement within errors. 

The better agreement between model 2 and AtomDB/APEC is expected, as both are based on collisional data from $R$-matrix calculations.  
The remaining small discrepancies might be caused by the slightly different atomic structures in AtomDB/APEC and SPEX. For instance,
the transition probabilities in AtomDB/APEC, taken from \citet{loch2006}, are higher 
than the SPEX values by 11\% and 6\% for the transitions of $2s^{2}2p^{5}3s$ $^3$P$_2$ to the ground and $2s^{2}2p^{5}3s$ $^1$P$_1$ to the 
ground, respectively.

Line ratios measured from three RGS objects, NGC~5044, NGC~3411, and the Fornax cluster, are found below the AtomDB/APEC curve.
These three data points are consistent with the model 1 curve within statistical errors. Physically, finding a line ratio below the theoretical curve
is unlikely, as there is no valid mechanism that could increase the 15~{\AA} or reduce the 17~{\AA} for a low-density
plasma in collisional ionization equilibrium. However, it is hard to conclude a significant preference among models 1, 2, and the AtomDB/APEC 
with the current RGS data.

The uncertainties in the optically thin line ratio would affect the random gas motion measured through the resonant scattering
phenomenon. This can be estimated based on the radiative transfer model 
of elliptical galaxies presented in \citet{anna2017}. As the optically thin ratio increases
by $\sim 12$\% from model 1 to AtomDB/APEC, the scattered photon flux would decrease by the same amount, and the inferred random motion
in the ICM would increase by a Mach number of $\geq 0.06$, or $\geq 12$\% of the mean gas motion reported in \citet{anna2017}. 

*/

\/*

\begin{landscape}

\longtab{
\begin{longtable}{l|c@{ }c@{ }c@{ }cc@{ }c@{ }c@{ }c@{ }cccccccccc}
\caption{\label{tab:rgsresult} Temperatures and abundances measured with the present and new plasma codes. For each object,
first and second lines are the results with the present and new codes, respectively. } \\
\hline
\hline
 \multicolumn{10}{c}{$3.4^{\prime}$ extraction regions}                                     \\
 \hline
            & \multicolumn{4}{c}{single-temperature} & \multicolumn{5}{c}{two-temperature$^{a}$}  \\         
\hline
cluster     & $T$ & O & Fe & C-stat & $T_1$ & $T_2$ & O & Fe & C-stat \\        

\endfirsthead
\caption{continued.}\\
\hline\hline
cluster     & $T$ & O & Fe & C-stat & $T_1$ & $T_2$ & O & Fe & C-stat \\        
\hline
\endhead
\hline
\endfoot

\hline
Abell~3526      & 1.60$\pm$0.01 & 0.40$\pm$0.01 & 0.69$\pm$0.01 & 2779/715 & 1.81$\pm$0.02 & 0.74$\pm$0.01 & 0.58$\pm$0.02 & 1.10$\pm$0.03 & 1297/713 \\
            & 1.55$\pm$0.01 & 0.39$\pm$0.01 & 0.59$\pm$0.01 & 2583/715 & 1.76$\pm$0.02 & 0.76$\pm$0.01 & 0.56$\pm$0.02 & 0.96$\pm$0.02 & 1226/713 \\
\hline
Fornax      & 0.95$\pm$0.01 & 0.50$\pm$0.03 & 0.29$\pm$0.02 & 1621/716 & 1.20$\pm$0.02 & 0.70$\pm$0.03 & 0.67$\pm$0.04 & 0.76$\pm$0.06 & 1280/714 \\
            & 0.98$\pm$0.01 & 0.40$\pm$0.03 & 0.28$\pm$0.02 & 1583/716 & 1.20$\pm$0.02 & 0.74$\pm$0.03 & 0.62$\pm$0.04 & 0.65$\pm$0.05 & 1276/714 \\
\hline
HCG~62      & 0.78$\pm$0.01 & 0.17$\pm$0.01 & 0.17$\pm$0.01 & 1220/714 & 1.10$\pm$0.02 & 0.65$\pm$0.02 & 0.24$\pm$0.02 & 0.38$\pm$0.02 &  996/712 \\ 
            & 0.80$\pm$0.01 & 0.18$\pm$0.01 & 0.16$\pm$0.01 & 1158/714 & 1.10$\pm$0.02 & 0.68$\pm$0.02 & 0.25$\pm$0.02 & 0.34$\pm$0.02 &  966/712 \\
\hline
M49        & 0.86$\pm$0.01 & 0.43$\pm$0.03 & 0.45$\pm$0.02 & 1198/712 & 1.11$\pm$0.03 & 0.72$\pm$0.02 & 0.56$\pm$0.04 & 0.85$\pm$0.06 & 1000/710 \\
            & 0.88$\pm$0.01 & 0.44$\pm$0.03 & 0.42$\pm$0.02 & 1125/712 & 1.12$\pm$0.03 & 0.75$\pm$0.02 & 0.56$\pm$0.05 & 0.76$\pm$0.05 &  961/710 \\
\hline
M86        & 0.70$\pm$0.01 & 0.23$\pm$0.02 & 0.19$\pm$0.01 &  984/710 & 1.01$\pm$0.03 & 0.62$\pm$0.02 & 0.27$\pm$0.02 & 0.30$\pm$0.03 &  920/708 \\  
            & 0.72$\pm$0.01 & 0.24$\pm$0.02 & 0.17$\pm$0.01 &  947/710 & 0.99$\pm$0.02 & 0.65$\pm$0.01 & 0.36$\pm$0.04 & 0.32$\pm$0.02 &  816/708 \\
\hline
NGC~1316    & 0.62$\pm$0.01 & 0.44$\pm$0.05 & 0.32$\pm$0.02 & 1362/713 & 1.20$\pm$0.10 & 0.59$\pm$0.01 & 0.58$\pm$0.06 & 0.53$\pm$0.05 & 1291/711 \\
            & 0.64$\pm$0.01 & 0.47$\pm$0.05 & 0.28$\pm$0.02 & 1325/713 & 1.20$\pm$0.10 & 0.61$\pm$0.01 & 0.66$\pm$0.04 & 0.50$\pm$0.01 & 1253/711 \\
\hline
NGC~1404    & 0.61$\pm$0.01 & 0.13$\pm$0.01 & 0.16$\pm$0.01 & 1045/713 & 0.63$\pm$0.04 & 0.50$\pm$0.10 & 0.13$\pm$0.01 & 0.17$\pm$0.01 & 1045/711 \\ 
            & 0.63$\pm$0.01 & 0.14$\pm$0.01 & 0.14$\pm$0.01 & 1010/713 & 0.65$\pm$0.02 & 0.48$\pm$0.09 & 0.14$\pm$0.01 & 0.14$\pm$0.01 & 1009/711 \\ 
\hline
NGC~3411    & 0.88$\pm$0.01 & 0.17$\pm$0.03 & 0.30$\pm$0.02 &  819/714 & 0.90$\pm$0.01 & 0.31$\pm$0.05 & 0.14$\pm$0.03 & 0.34$\pm$0.02 &  810/712 \\
            & 0.89$\pm$0.01 & 0.18$\pm$0.03 & 0.28$\pm$0.01 &  812/714 & 0.91$\pm$0.01 & 0.29$\pm$0.07 & 0.16$\pm$0.02 & 0.30$\pm$0.02 &  808/712 \\
\hline
NGC~4325    & 0.90$\pm$0.02 & 0.10$\pm$0.02 & 0.20$\pm$0.02 &  865/715 & 0.96$\pm$0.02 & 0.33$\pm$0.04 & 0.08$\pm$0.02 & 0.29$\pm$0.03 &  842/710 \\
            & 0.92$\pm$0.02 & 0.11$\pm$0.02 & 0.19$\pm$0.02 &  856/715 & 0.97$\pm$0.03 & 0.34$\pm$0.06 & 0.10$\pm$0.02 & 0.24$\pm$0.04 &  833/710 \\
\hline
NGC~4374    & 0.64$\pm$0.01 & 0.22$\pm$0.02 & 0.17$\pm$0.01 &  845/710 & 1.04$\pm$0.07 & 0.61$\pm$0.02 & 0.23$\pm$0.02 & 0.23$\pm$0.02 &  810/708 \\
            & 0.66$\pm$0.01 & 0.22$\pm$0.02 & 0.14$\pm$0.01 &  824/710 & 1.02$\pm$0.06 & 0.62$\pm$0.02 & 0.24$\pm$0.02 & 0.20$\pm$0.02 &  794/708 \\
\hline
NGC~4636    & 0.65$\pm$0.01 & 0.29$\pm$0.02 & 0.38$\pm$0.01 & 1174/718 & 0.88$\pm$0.04 & 0.58$\pm$0.02 & 0.29$\pm$0.02 & 0.44$\pm$0.02 & 1080/716 \\
            & 0.67$\pm$0.01 & 0.31$\pm$0.02 & 0.33$\pm$0.01 & 1071/718 & 0.84$\pm$0.04 & 0.59$\pm$0.02 & 0.30$\pm$0.02 & 0.38$\pm$0.02 &  984/716 \\
\hline
NGC~4649    & 0.81$\pm$0.01 & 0.28$\pm$0.02 & 0.35$\pm$0.02 &  931/712 &               &               &               &               &          \\
            & 0.83$\pm$0.01 & 0.28$\pm$0.02 & 0.31$\pm$0.01 &  906/712 &               &               &               &               &          \\
\hline
NGC~5044    & 0.84$\pm$0.01 & 0.35$\pm$0.01 & 0.38$\pm$0.01 & 1401/713 & 1.00$\pm$0.02 & 0.69$\pm$0.02 & 0.41$\pm$0.02 & 0.59$\pm$0.01 & 1071/711 \\
            & 0.86$\pm$0.01 & 0.36$\pm$0.01 & 0.37$\pm$0.01 & 1205/713 & 1.01$\pm$0.02 & 0.74$\pm$0.02 & 0.42$\pm$0.02 & 0.53$\pm$0.01 &  993/711 \\
\hline
NGC~5813    & 0.66$\pm$0.01 & 0.35$\pm$0.02 & 0.53$\pm$0.01 & 1434/715 & 0.73$\pm$0.01 & 0.45$\pm$0.03 & 0.34$\pm$0.02 & 0.62$\pm$0.02 & 1344/713 \\
            & 0.68$\pm$0.01 & 0.37$\pm$0.02 & 0.48$\pm$0.01 & 1140/715 & 0.73$\pm$0.01 & 0.42$\pm$0.01 & 0.34$\pm$0.02 & 0.54$\pm$0.02 & 1073/713 \\
\hline
NGC~5846    & 0.64$\pm$0.01 & 0.26$\pm$0.01 & 0.28$\pm$0.01 & 1067/714 & 0.81$\pm$0.02 & 0.53$\pm$0.02 & 0.26$\pm$0.01 & 0.34$\pm$0.01 &  984/712  \\
            & 0.66$\pm$0.01 & 0.27$\pm$0.01 & 0.25$\pm$0.01 &  916/714 & 0.79$\pm$0.02 & 0.52$\pm$0.03 & 0.27$\pm$0.01 & 0.31$\pm$0.01 &  841/712  \\
\hline
 \multicolumn{10}{c}{$0.8^{\prime}$ extraction regions}                                     \\
 \hline
            & \multicolumn{4}{c}{single-temperature} & \multicolumn{5}{c}{two-temperature$^{a}$}  \\         
\hline
cluster     & $T$ & O & Fe & C-stat & $T_1$ & $T_2$ & O & Fe & C-stat \\        
\hline

Abell~3526      & 1.33$\pm$0.01 & 0.39$\pm$0.01 & 0.47$\pm$0.01 & 2571/715 & 1.62$\pm$0.02 & 0.72$\pm$0.01 & 0.63$\pm$0.02 & 1.08$\pm$0.03 &  985/713 \\
            & 1.33$\pm$0.01 & 0.39$\pm$0.01 & 0.44$\pm$0.01 & 2389/715 & 1.59$\pm$0.02 & 0.73$\pm$0.01 & 0.62$\pm$0.02 & 0.94$\pm$0.03 &  933/713 \\
\hline  
Fornax      & 0.87$\pm$0.01 & 0.35$\pm$0.03 & 0.39$\pm$0.02 &  984/716 & 1.22$\pm$0.11 & 0.79$\pm$0.02 & 0.47$\pm$0.04 & 0.70$\pm$0.05 &  845/714 \\
            & 0.89$\pm$0.01 & 0.36$\pm$0.03 & 0.37$\pm$0.02 &  958/716 & 1.32$\pm$0.12 & 0.84$\pm$0.02 & 0.49$\pm$0.04 & 0.66$\pm$0.05 &  842/714 \\
\hline  
HCG~62      & 0.74$\pm$0.01 & 0.24$\pm$0.02 & 0.31$\pm$0.02 &  901/714 & 1.03$\pm$0.04 & 0.65$\pm$0.02 & 0.30$\pm$0.03 & 0.51$\pm$0.03 &  814/712 \\
            & 0.77$\pm$0.01 & 0.26$\pm$0.03 & 0.29$\pm$0.01 &  853/714 & 1.03$\pm$0.04 & 0.67$\pm$0.03 & 0.31$\pm$0.03 & 0.45$\pm$0.03 &  782/712 \\
\hline
M49        & 0.79$\pm$0.01 & 0.41$\pm$0.04 & 0.48$\pm$0.03 &  885/712 & [$\equiv$1.11]$^{b}$& 0.77$\pm$0.01 & 0.78$\pm$0.09 & 1.10$\pm$0.10 &  804/710 \\
            & 0.81$\pm$0.01 & 0.42$\pm$0.04 & 0.44$\pm$0.02 &  835/712 & [$\equiv$1.12]$^{b}$& 0.78$\pm$0.01 & 0.79$\pm$0.09 & 1.03$\pm$0.10 &  755/710 \\
\hline
M86        & 0.69$\pm$0.01 & 0.29$\pm$0.04 & 0.23$\pm$0.02 &  850/710 & 1.03$\pm$0.02 & 0.65$\pm$0.01 & 0.35$\pm$0.03 & 0.38$\pm$0.02 &  833/708 \\
            & 0.71$\pm$0.01 & 0.32$\pm$0.04 & 0.21$\pm$0.02 &  830/710 & 1.00$\pm$0.02 & 0.65$\pm$0.01 & 0.36$\pm$0.03 & 0.32$\pm$0.02 &  816/708 \\
\hline
NGC~1316    & 0.62$\pm$0.01 & 0.44$\pm$0.05 & 0.35$\pm$0.03 &  997/713 & [$\equiv$1.20]$^{b}$& 0.58$\pm$0.01 & 0.59$\pm$0.07 & 0.59$\pm$0.09 &  942/711 \\
            & 0.64$\pm$0.01 & 0.47$\pm$0.05 & 0.31$\pm$0.02 &  965/713 & [$\equiv$1.20]$^{b}$& 0.60$\pm$0.01 & 0.63$\pm$0.08 & 0.53$\pm$0.05 &  913/711 \\
\hline
NGC~1404    & 0.61$\pm$0.01 & 0.17$\pm$0.02 & 0.21$\pm$0.01 &  945/713 &               &               &               &               &          \\ 
            & 0.63$\pm$0.01 & 0.18$\pm$0.02 & 0.18$\pm$0.01 &  916/713 &               &               &               &               &          \\
\hline
NGC~3411    & 0.88$\pm$0.01 & 0.23$\pm$0.05 & 0.49$\pm$0.04 &  788/714 &               &               &               &               &          \\
            & 0.90$\pm$0.01 & 0.24$\pm$0.05 & 0.46$\pm$0.04 &  782/714 &               &               &               &               &          \\
\hline
NGC~4325    & 0.85$\pm$0.02 & 0.18$\pm$0.04 & 0.33$\pm$0.03 &  781/712 & 0.92$\pm$0.02 & 0.40$\pm$0.09 & 0.17$\pm$0.04 & 0.44$\pm$0.04 &  767/710 \\
            & 0.87$\pm$0.02 & 0.19$\pm$0.04 & 0.32$\pm$0.03 &  772/712 & 0.92$\pm$0.02 & 0.37$\pm$0.09 & 0.17$\pm$0.04 & 0.39$\pm$0.04 &  762/710 \\
\hline
NGC~4374    & 0.61$\pm$0.01 & 0.30$\pm$0.03 & 0.27$\pm$0.01 &  828/710 & 1.00$\pm$0.09 & 0.58$\pm$0.02 & 0.32$\pm$0.04 & 0.33$\pm$0.03 &  813/708 \\ 
            & 0.64$\pm$0.01 & 0.32$\pm$0.03 & 0.23$\pm$0.01 &  809/710 & 0.97$\pm$0.10 & 0.60$\pm$0.03 & 0.33$\pm$0.04 & 0.28$\pm$0.03 &  794/708 \\
\hline
NGC~4636    & 0.60$\pm$0.01 & 0.37$\pm$0.03 & 0.47$\pm$0.03 & 1009/718 & 0.83$\pm$0.04 & 0.54$\pm$0.02 & 0.35$\pm$0.03 & 0.53$\pm$0.03 &  932/716 \\
            & 0.62$\pm$0.01 & 0.38$\pm$0.03 & 0.41$\pm$0.02 &  929/718 & 0.81$\pm$0.04 & 0.55$\pm$0.02 & 0.36$\pm$0.03 & 0.45$\pm$0.03 &  853/716 \\
\hline
NGC~4649    & 0.81$\pm$0.01 & 0.40$\pm$0.03 & 0.55$\pm$0.03 &  883/712 &               &               &               &               &          \\
            & 0.83$\pm$0.01 & 0.40$\pm$0.03 & 0.51$\pm$0.02 &  852/712 &               &               &               &               &          \\ 
\hline
NGC~5044    & 0.79$\pm$0.01 & 0.39$\pm$0.02 & 0.43$\pm$0.01 &  990/713 &               &               &               &               &          \\ 
            & 0.81$\pm$0.01 & 0.41$\pm$0.02 & 0.41$\pm$0.01 &  918/713 &               &               &               &               &          \\ 
\hline
NGC~5813    & 0.65$\pm$0.01 & 0.42$\pm$0.02 & 0.62$\pm$0.02 & 1094/715 &               &               &               &               &          \\ 
            & 0.67$\pm$0.01 & 0.44$\pm$0.02 & 0.55$\pm$0.02 &  918/715 &               &               &               &               &          \\ 
\hline
NGC~5846    & 0.63$\pm$0.01 & 0.34$\pm$0.02 & 0.38$\pm$0.02 &  908/714 & 0.77$\pm$0.03 & 0.51$\pm$0.03 & 0.33$\pm$0.03 & 0.45$\pm$0.02 &  842/712 \\
            & 0.65$\pm$0.01 & 0.36$\pm$0.02 & 0.34$\pm$0.01 &  812/714 & 0.76$\pm$0.03 & 0.48$\pm$0.03 & 0.35$\pm$0.03 & 0.40$\pm$0.02 &  776/712 \\
\hline

\end{longtable}
\begin{tablenotes}
\item[$(a)$] $a:$ Columns left empty when the second component is not required from the fits.
\item[$(b)$] $b:$ Fixed to the values obtained in the $3.4^{\prime}$ extraction regions.
\end{tablenotes}
}
\end{landscape}

*/

\section{Ending remarks}

By carrying out a large-scale theoretical calculation of the electron impact
on ions of \ion{Fe}{XVII} to \ion{Fe}{XXIV}, we present a set of new atomic data 
for modeling the Fe-L complex. The calculation includes a large set of atomic levels
for each ion, allowing full configuration interaction and all kinds of resonant processes. 
The resonant excitation and the dielectronic recombination are found to affect 
strongly a significant portion of the major transitions. We present a set of detailed 
comparisons of the new calculation with available $R$-matrix results, on both
the collisional rates and the model spectra based on the line formation calculations. 
It shows that the two calculations agree within $20$\% on most of the main transitions.
The comparison will be fed into the
prioritization of the future laboratory benchmarks on the Fe-L modeling.

%More discrepancies are seen for the weaker transitions and satellite lines, in particular
%for \ion{Fe}{XVIII}, \ion{Fe}{XIX}, and \ion{Fe}{XX}. 

The current SPEX code includes mostly non-resonant atomic calculations.
The fact that many previous RGS results on elliptical galaxies and galaxy clusters are 
based on a non-resonant database is worrying. To assess the possible bias, we apply
the new FAC calculation with complete resonances to a RGS sample of 15 cool-core ellipticals and clusters. 
%By applying the new FAC calculation to a RGS sample of 
%15 elliptical galaxies and galaxy clusters with temperatures of $0.6-1.5$ keV, 
We find that the Fe abundances measured with the current SPEX v3.04 code are on average
biased high by 12\%. The O/Fe abundance ratio, which is widely used for assessing the 
population of the enriching supernovae,
is underestimated by a mean value of 16\%. Furthermore, the Fe abundances measured with the $R$-matrix model
show discrepancies of $\leq 10$\% from the values with the FAC model. Current data cannot decisively distinguish 
between the FAC and the $R$-matrix models, therefore the 10\% abundance difference
has to be treated as systematic uncertainties. 
%Similar uncertainties on the 
%abundances are also seen in the DEM fit of the Capella HETG spectrum. 

To update the atomic database in a plasma code (such as SPEX), it is ideal to take the best from
 the $R$-matrix and the FAC calculations. In theory, the accuracy of $R$-matrix data
 might be considered superior to that of the FAC calculation. The $R$-matrix data should be implemented on
 the low-lying levels, which form the main X-ray transitions. For the high levels,
as the $R$-matrix data gradually becomes sparse, the new FAC calculation with isolated resonances can be implemented
as a valid approximation. This would form a recommended database used in most of the analysis. On the other hand,
it might be desirable to keep a second database with the new FAC calculations for both the low and high levels. Since the
$R$-matrix and FAC calculations might represent two ends of the theoretical space, comparing the fits with 
the recommended and the second databases might directly reflect the systematic atomic uncertainties on the source parameters.

The next step of the Fe-L work will be twofold. First, the test with astrophysical objects with existing
observatories will be continued. As shown in the test with the RGS data, benchmarks using astrophysical objects require
not only a compatible atomic database, but also a proper analysis technique for modeling out the astrophysical effects. Second, 
we will put forward a dedicated benchmark with ground-based laboratory experiments using electron beam ion trap devices, where 
plasma in a Maxwellian distribution can be simulated. By checking the consistency between the models and
the astrophysical/experimental spectra for each visible Fe-L transition, we will
identify the potential areas where the theoretical calculations can be further improved. Some iterations of 
such work will be needed to ensure that the atomic codes are ready for the future high resolution X-ray spectra
obtained with {\it XRISM} and {\it Athena}.

\begin{acknowledgements}

L.G. is supported by the RIKEN Special Postdoctoral Researcher Program.
SRON is supported financially by NWO, the Netherlands Organization for
Scientific Research. A. S. is supported by the Women In Science Excel 
(WISE) programme of the Netherlands Organisation for Scientific Research 
(NWO), and acknowledges the MEXT World Premier Research Center Initiative 
(WPI) and the Kavli IPMU for the continued hospitality.

\end{acknowledgements}

\bibliographystyle{aa}
\bibliography{main}

\newpage

\begin{appendix}

\section{Density effects}
\label{sec:density}

\begin{figure*}[!htbp]
\centering
\resizebox{0.9\hsize}{!}{\includegraphics[angle=0]{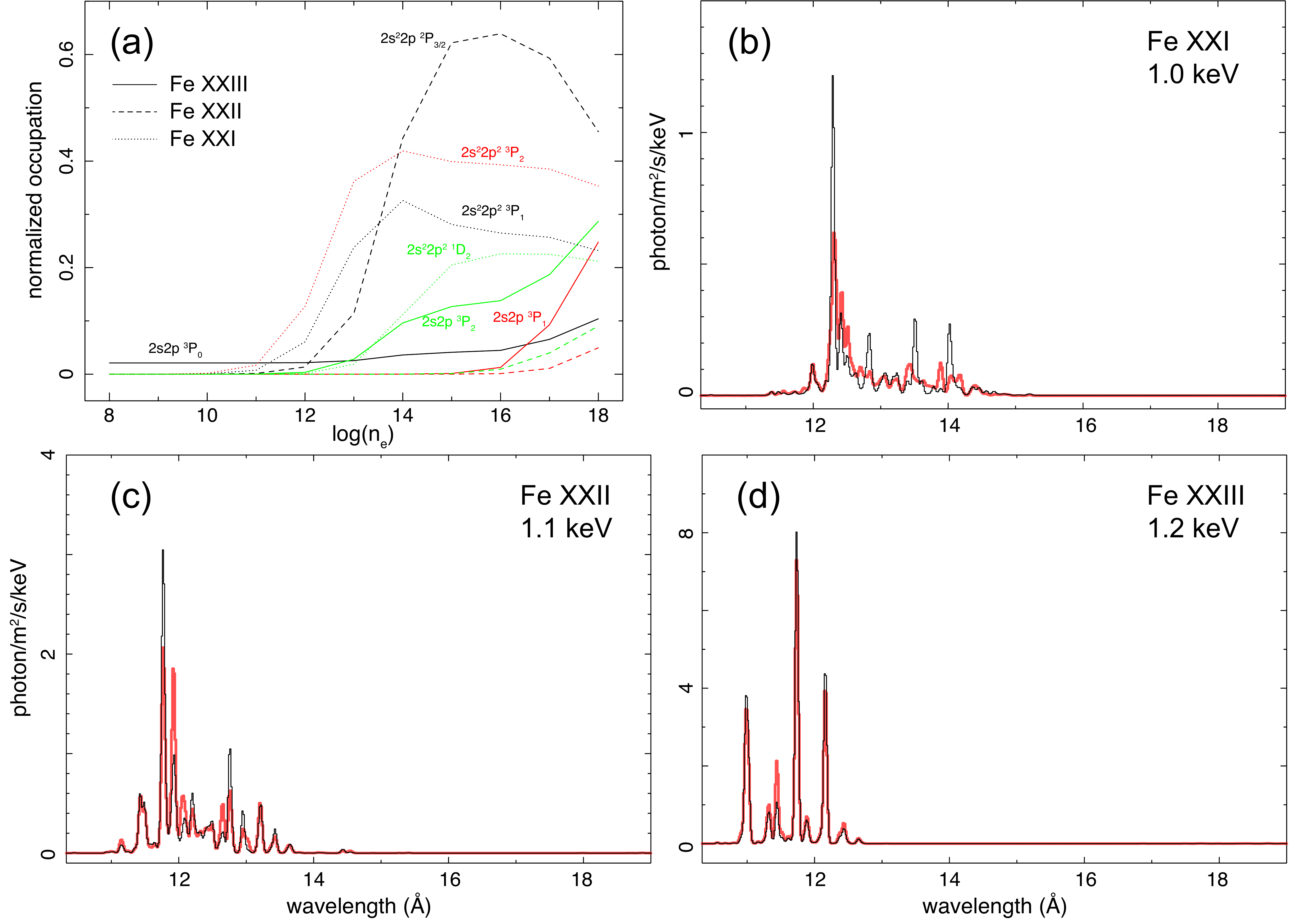}}
\caption{(a) The normalized population of the three low-lying excited levels of C-like, B-like, and Be-like 
Fe, as a function of electron density. The population of the ground states is not plotted. The density
effect on the model spectra for the C-like, B-like, and Be-like Fe is shown in (b), (c), and (d), respectively.
The spectra of low-density (1 cm$^{-3}$) and intermediate density ($10^{14}$ cm$^{-3}$) are plotted in black and red. The 
temperatures are the same as in Fig.~\ref{fig:facspec}. }
\label{fig:population}
\end{figure*}

For a low-density plasma, excited levels are quickly depleted by spontaneous 
cascade, so that only the ground state is significantly populated. At a high electron density, 
the cascade might be interrupted by collision with electrons. Some of the low-lying
levels become thus populated. As a result, the population of the ground state decreases,
and the related spectral features, e.g., lines from ground excitation, become weaker.
On the other side, the transitions from the excited states become substantially 
more important \citep{mao2017}. 

Figure~\ref{fig:population} illustrates the density-dependent population
of several low-lying excited levels for the C-like, B-like, and Be-like Fe. The calculation
is done with SPEX, which incorporates the new atomic data obtained in this work. For 
an electron density lower than $10^{10}$ cm$^{-3}$, the occupations of these levels
are negligible, except for the metastable 2$s$2$p$ $^3$P$_{0}$ level, which is populated  
even at low density due to the narrow de-excitation channel. For the selected low-lying
levels shown in the figure, the population rises as the density increases from $10^{12}$ cm$^{-3}$
to $10^{14}$ cm$^{-3}$. At a higher density, the relative level population evolves 
towards the standard Boltzman distribution, as the excited states would eventually be in a collisional
local thermodynamic equilibrium (LTE). 

The same exercise has been done for the other Fe-L species. For the astrophysical coronal/nebular 
($< 10^{14}$ cm$^{-3}$) plasma, the density effect is most 
significant at the three low-lying excited levels for the Fe-L. For the three levels, we calculate the 
transition rates of direct excitation, resonant
excitation, and dielectronic recombination, in a same way as those from the ground states (\S~\ref{sec:method}). 

It should be noted that there could be more metastable levels above the three low-lying levels included
in the current calculation. \citet{badnell2006} included 6 low-lying excited levels for O-like Fe and Be-like Fe, 
8 for N-like and B-like, and 12 for C-like, as the metastable parent levels used in the radiative recombination calculation. 
The extra metastable levels would become sensitive for the condition 
of higher density ($> 10^{14}$ cm$^{-3}$). We plan to include all the metastable levels in a follow-up calculation.

Figure~\ref{fig:population} shows the model spectra based on the above data, for 
the C-like, B-like, and Be-like Fe at a low density and an intermediate density of $10^{14}$ cm$^{-3}$. 
The temperature is set to the value of peak ion concentration in equilibrium. It can be seen
that the dominant lines of these ions become weaker at high density, probably because these lines
originate from the excitation of the ground states, which have a decreasing population at high
density. Some of the satellite lines become stronger, as the low-lying levels contribute significantly
to the formation of these lines.

\section{Resonant excitation: consistency check with previous results}
\label{sec:re_check}

Following \S~\ref{sec:res_result}, we compare the electron-impact collision
data obtained from the current FAC calculation with those from
previous distorted wave and $R$-matrix works.

\begin{table*}[!htbp]
\centering
\caption{ Partial resonance rates for the 2$p^5$3$s$ levels of \ion{Fe}{XVII}}
\label{tab:partial}
\begin{threeparttable}
\begin{tabular}{lcccccccccccccccc}
\hline
Final state & calculation & rate at 0.2~keV$^{a}$ & rate at 0.4~keV$^{b}$ & rate at 1.0~keV \\ 
            &             &                 & $10^{-13}$ cm$^{3}$ s$^{-1}$ &    \\
\hline 
2$s^2$2$p^5$3$s$ $^3$P$_2$ & \citet{smith1985} & 50.0   & $-$    & $-$ \\
                         & \citet{gold1989}  & 13.0   & 29.9   & $-$ \\
                         & \citet{chen1989}  & 12.2   & 30.2   & 22.8 \\
                         & this work           & 19.0   & 44.0   & 34.5 \\
\hline
2$s^2$2$p^5$3$s$ $^1$P$_1$ & \citet{smith1985} & 48.0   & $-$    & $-$ \\
                         & \citet{gold1989}  & 16.1   & 37.6   & $-$ \\
                         & \citet{chen1989}  & 14.4   & 36.9   & 27.9 \\
                         & this work           & 17.2   & 40.6   & 32.2 \\
\hline
2$s^2$2$p^5$3$s$ $^3$P$_0$ & \citet{smith1985} & 11.0   & $-$    & $-$ \\
                         & \citet{gold1989}  & 2.5    & 6.0    & $-$  \\
                         & \citet{chen1989}  & 1.9    & 5.0    & 3.8  \\
                         & this work           & 2.9    & 6.9    & 5.6  \\
\hline
2$s^2$2$p^5$3$s$ $^3$P$_1$ & \citet{smith1985} & 38.0   & $-$    & $-$ \\
                         & \citet{gold1989}  & 15.1   & 36.7   & $-$ \\
                         & \citet{chen1989}  & 14.2   & 37.1   & 28.5 \\
                         & \citet{chen2002}  & 22.7   & $-$    & $-$ \\  
                         & this work           & 16.0   & 38.9   & 31.4 \\
\hline
All 2$s^2$2$p^5$3$s$ states & \citet{smith1985} & 147.0 & $-$    & $-$ \\
                          & \citet{gold1989}  & 46.7  & 110.1  & $-$ \\
                          & \citet{chen1989}  & 42.7  & 109.2  & 83.0\\
                          & \citet{chen2002}  & 70.4  & $-$    & $-$ \\
                          & this work           & 55.1  & 130.4  & 103.7\\
\hline
\end{tabular}
\begin{tablenotes}
\item[$(a)$] Values from \citet{smith1985} are calculated at 217 eV. 
\item[$(b)$] Values from \citet{chen1989} are calculated at 0.5~keV. 
\end{tablenotes}
\end{threeparttable}
\end{table*}

\subsection{Comparing with classic \ion{Fe}{XVII} resonance calculations}

In Table~\ref{tab:partial}, the resonance rates of the \ion{Fe}{XVII} ion from the present work are compared with the results in literature. The
values from \citet{smith1985}, \citet{gold1989}, \citet{chen1989}, and \citet{chen2002} are based on the semi-relativistic Hartree-Fock method, a relativistic
parametric potential method, the multi-configuration Dirac-Fock approach, and a Breit-Pauli $R$-matrix method with a 89-level expansion, respectively. 
The rate coefficients reported in \citet{smith1985}
are apparently higher than the others. \citet{chen1989} suggested that this is partially explained by the incomplete autoionization channels included in \citet{smith1985}.
The present calculation gives higher rates than the those of \citet{gold1989} and \citet{chen1989} using a similar technique, but lower than the Breit-Pauli $R$-matrix results. 
The discrepancies between the \citet{chen2002} values and our results at 0.2~keV are 42\% on the partial rate to 2$s^2$2$p^5$3$s$ $^3$P$_1$, and 28\% on the total rate.

\begin{figure*}[!htbp]
\centering
\resizebox{0.9\hsize}{!}{\includegraphics[angle=0]{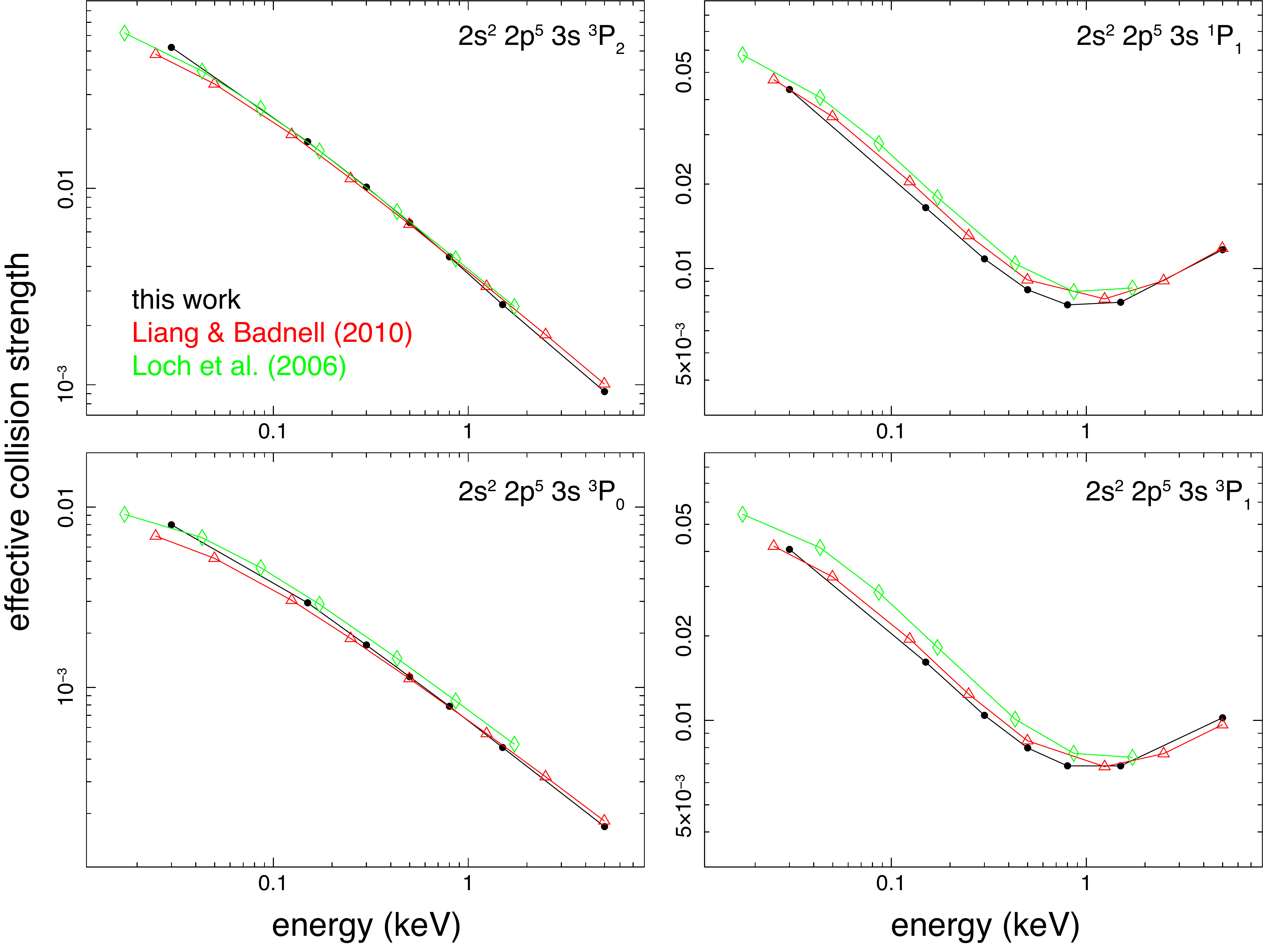}}
\caption{Maxwellian-averaged collision strengths for four main \ion{Fe}{XVII} transitions. Red and green curves
are the $R$-matrix data 
taken from \citet{liang2010} and \citet{loch2006}. }
\label{fig:rmatrix1}
\end{figure*}

\begin{figure*}[!htbp]
\centering
\resizebox{0.9\hsize}{!}{\includegraphics[angle=0]{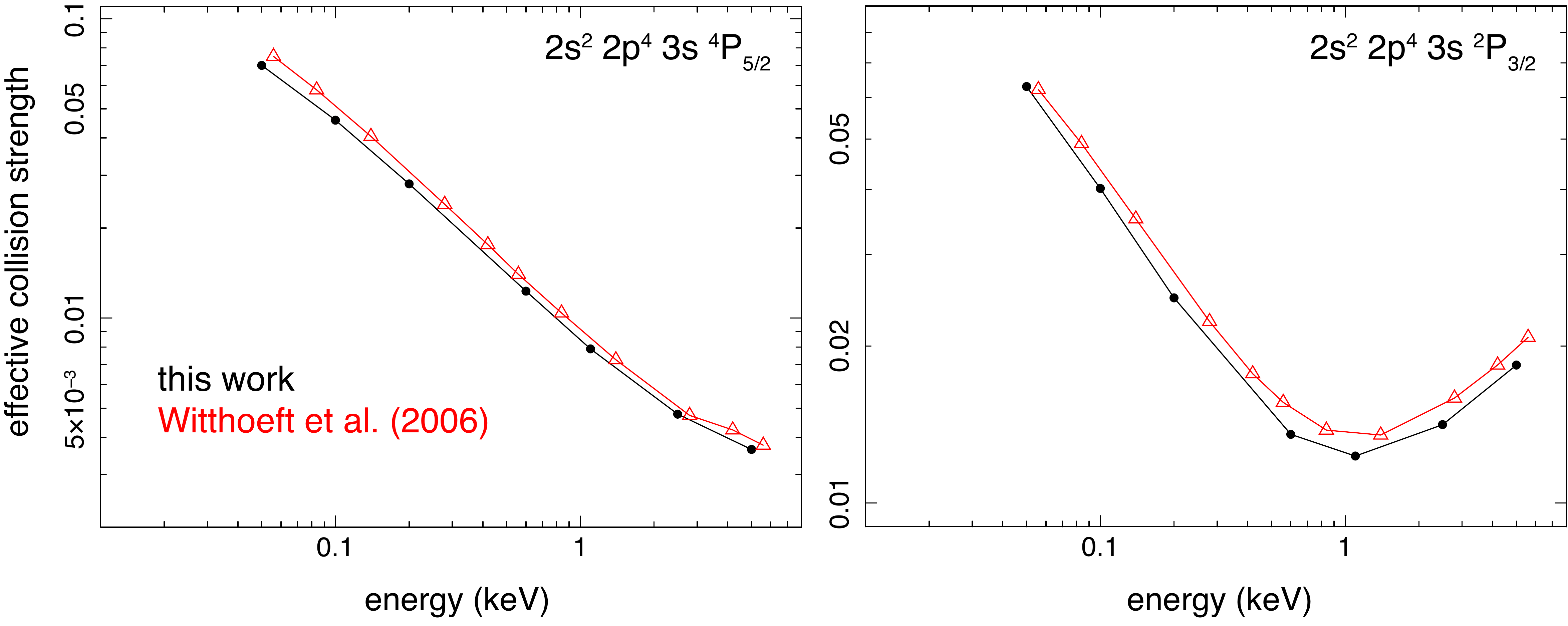}}
\caption{Same as Fig.~\ref{fig:rmatrix1} but for two \ion{Fe}{XVIII} lines. The $R$-matrix data from \citet{witt2006} are shown in red.}
\label{fig:rmatrix2}
\end{figure*}

\begin{figure*}[!htbp]
\centering
\resizebox{0.9\hsize}{!}{\includegraphics[angle=0]{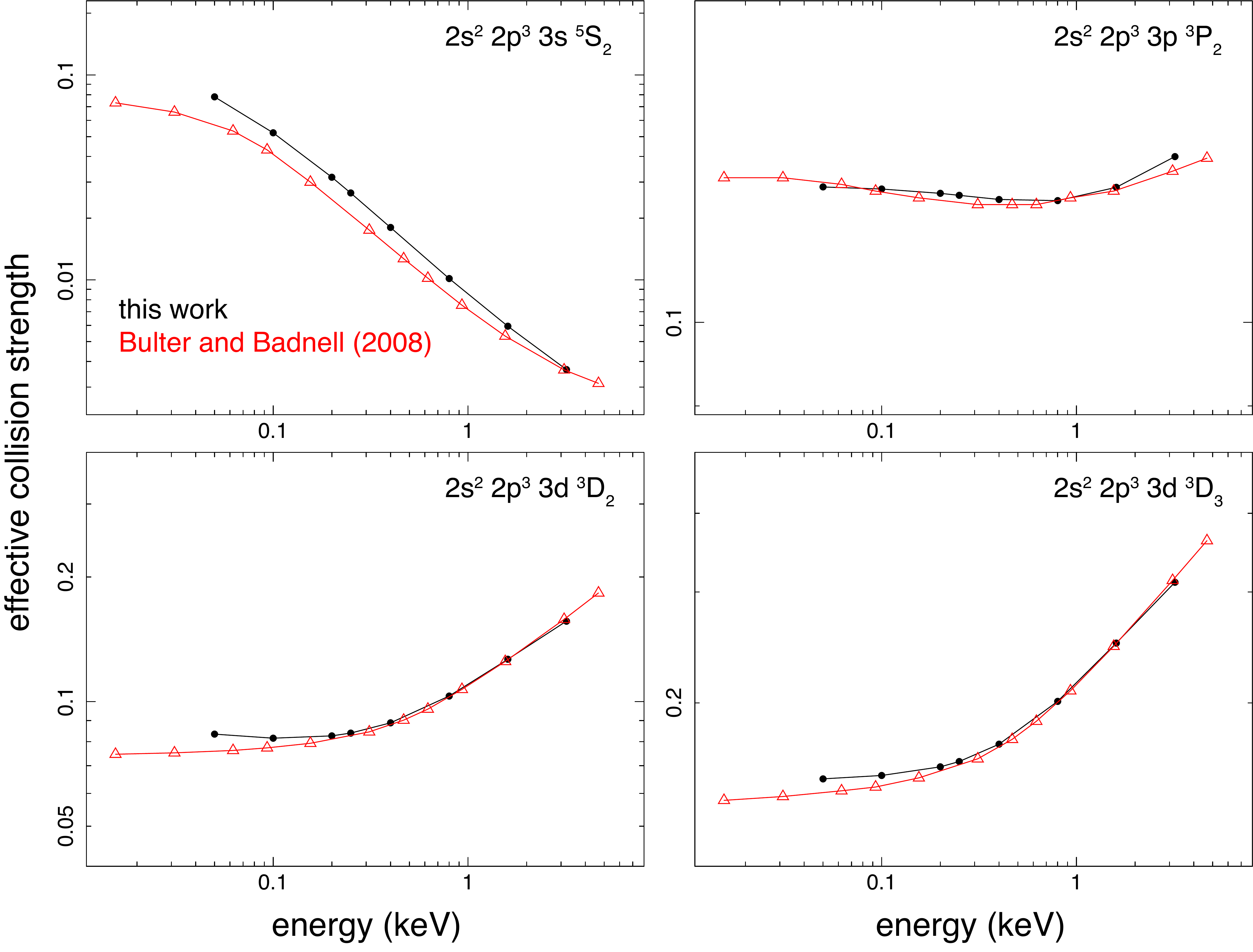}}
\caption{Same as Fig.~\ref{fig:rmatrix1} but for four \ion{Fe}{XIX} lines. The $R$-matrix data from \citet{butler2008} are shown in red.}
\label{fig:rmatrix22}
\end{figure*}

\begin{figure*}[!htbp]
\centering
\resizebox{0.9\hsize}{!}{\includegraphics[angle=0]{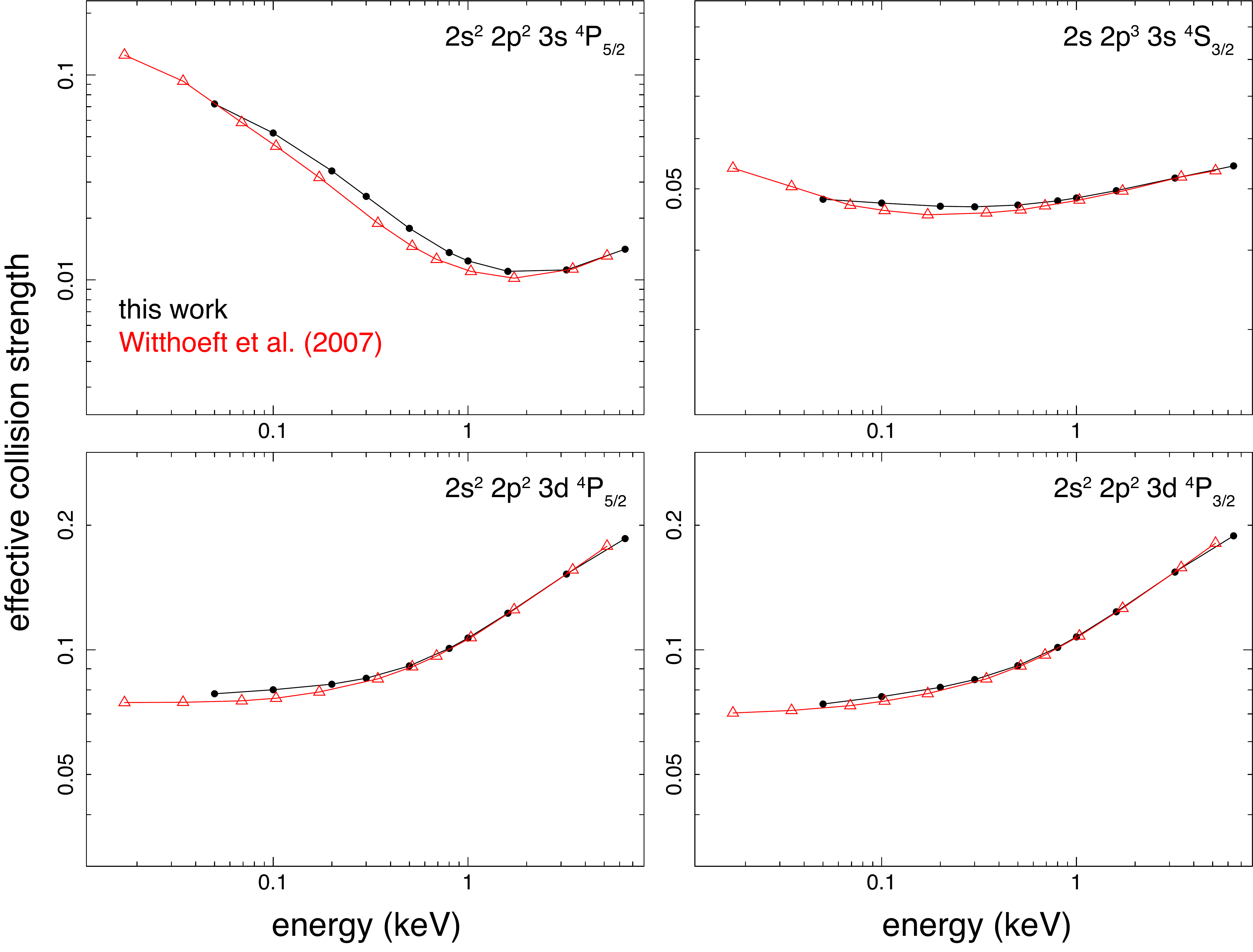}}
\caption{Same as Fig.~\ref{fig:rmatrix1} but for four \ion{Fe}{XX} lines. The $R$-matrix data from \citet{witt2007} are shown in red.}
\label{fig:rmatrix23}
\end{figure*}

\begin{figure*}[!htbp]
\centering
\resizebox{0.9\hsize}{!}{\includegraphics[angle=0]{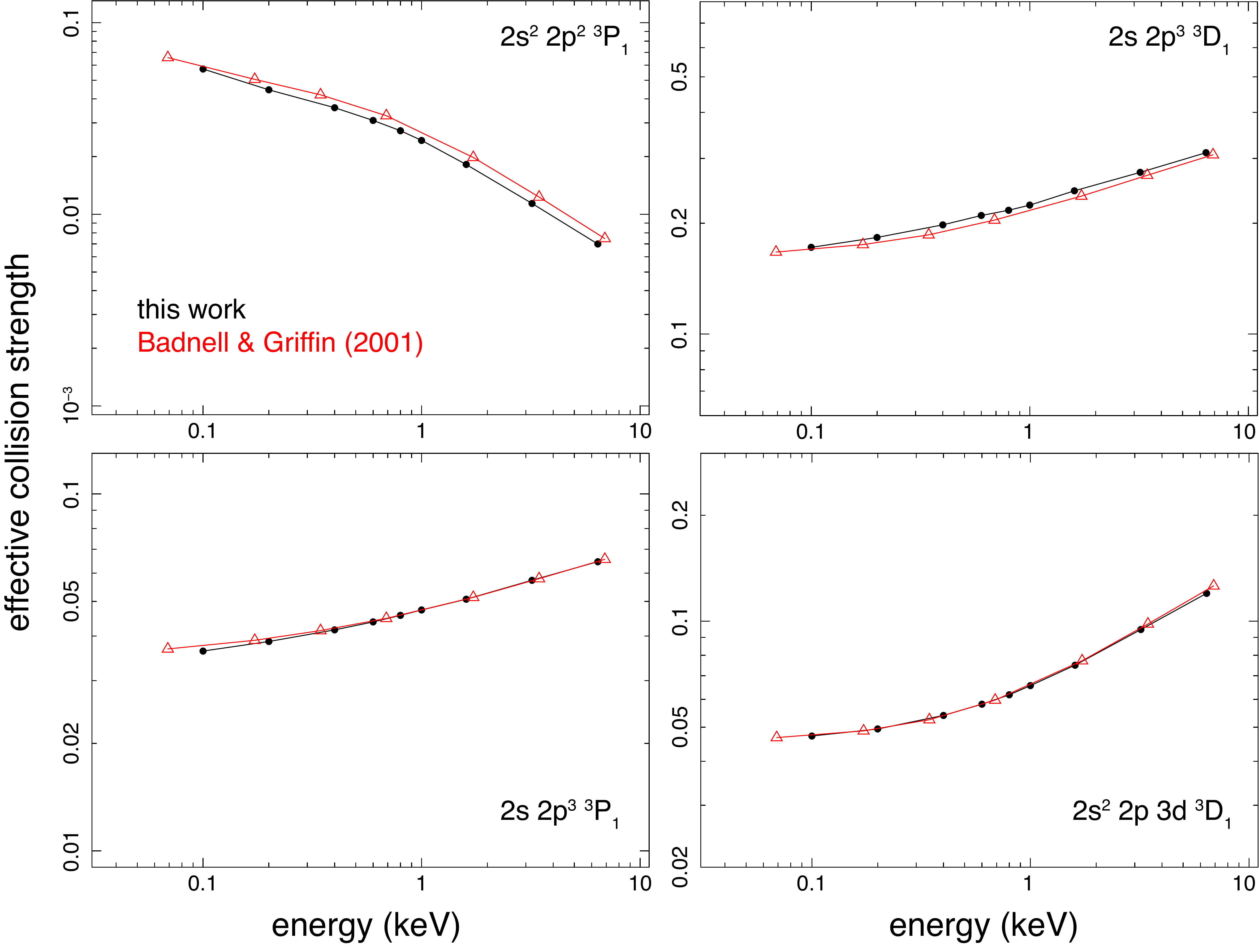}}
\caption{Same as Fig.~\ref{fig:rmatrix1} but for four \ion{Fe}{XXI} lines. The $R$-matrix data from \citet{badnell2001} are plot in red.}
\label{fig:rmatrix3}
\end{figure*}

\begin{figure*}[!htbp]
\centering
\resizebox{0.9\hsize}{!}{\includegraphics[angle=0]{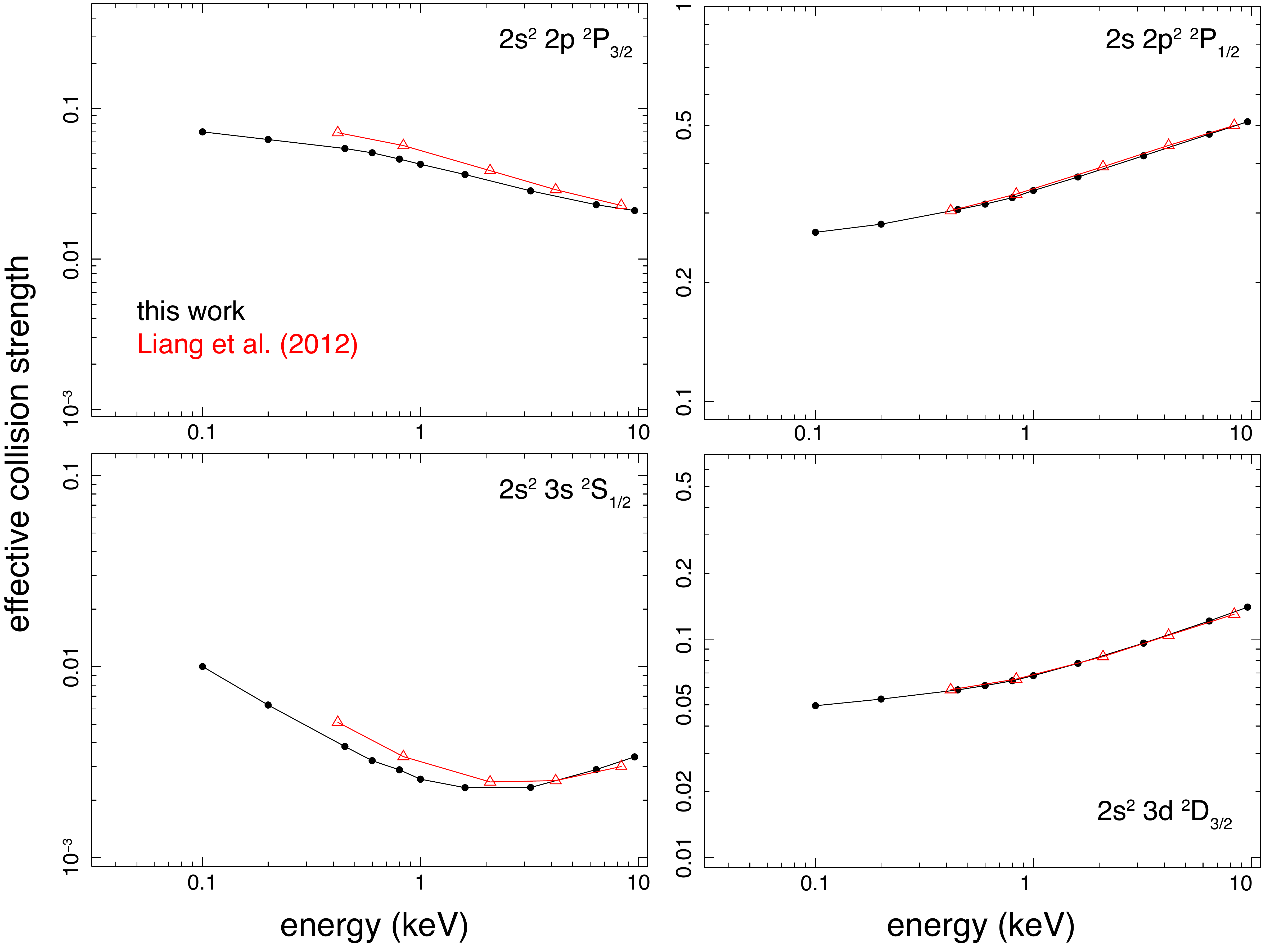}}
\caption{Same as Fig.~\ref{fig:rmatrix1} but for four \ion{Fe}{XXII} lines. The $R$-matrix data from \citet{liang2012} are plot in red.}
\label{fig:rmatrix4}
\end{figure*}

\begin{figure*}[!htbp]
\centering
\resizebox{0.9\hsize}{!}{\includegraphics[angle=0]{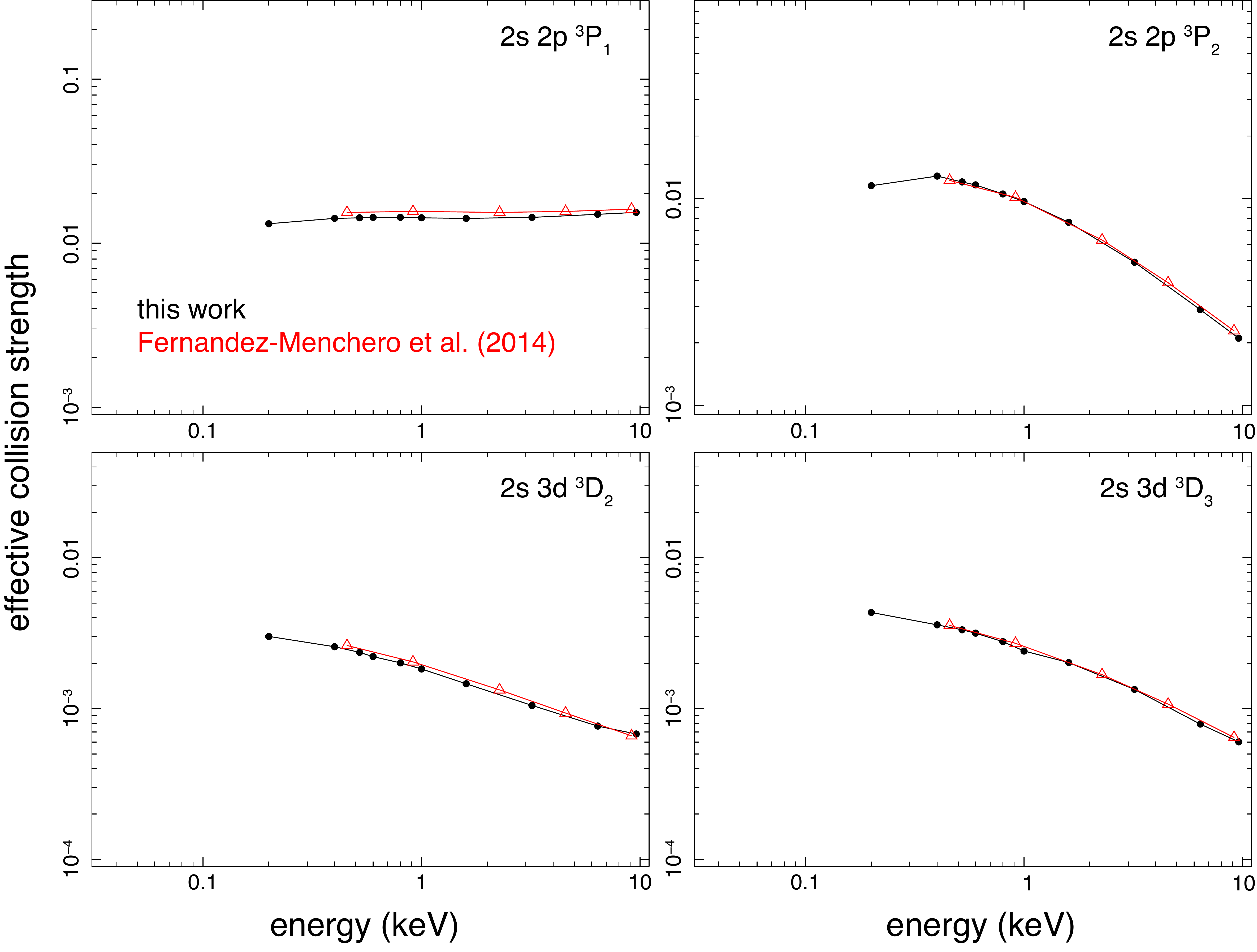}}
\caption{Same as Fig.~\ref{fig:rmatrix1} but for four \ion{Fe}{XXIII} lines. The $R$-matrix data from \citet{fern2014} are shown in red.}
\label{fig:rmatrix5}
\end{figure*}

\begin{figure*}[!htbp]
\centering
\resizebox{0.9\hsize}{!}{\includegraphics[angle=0]{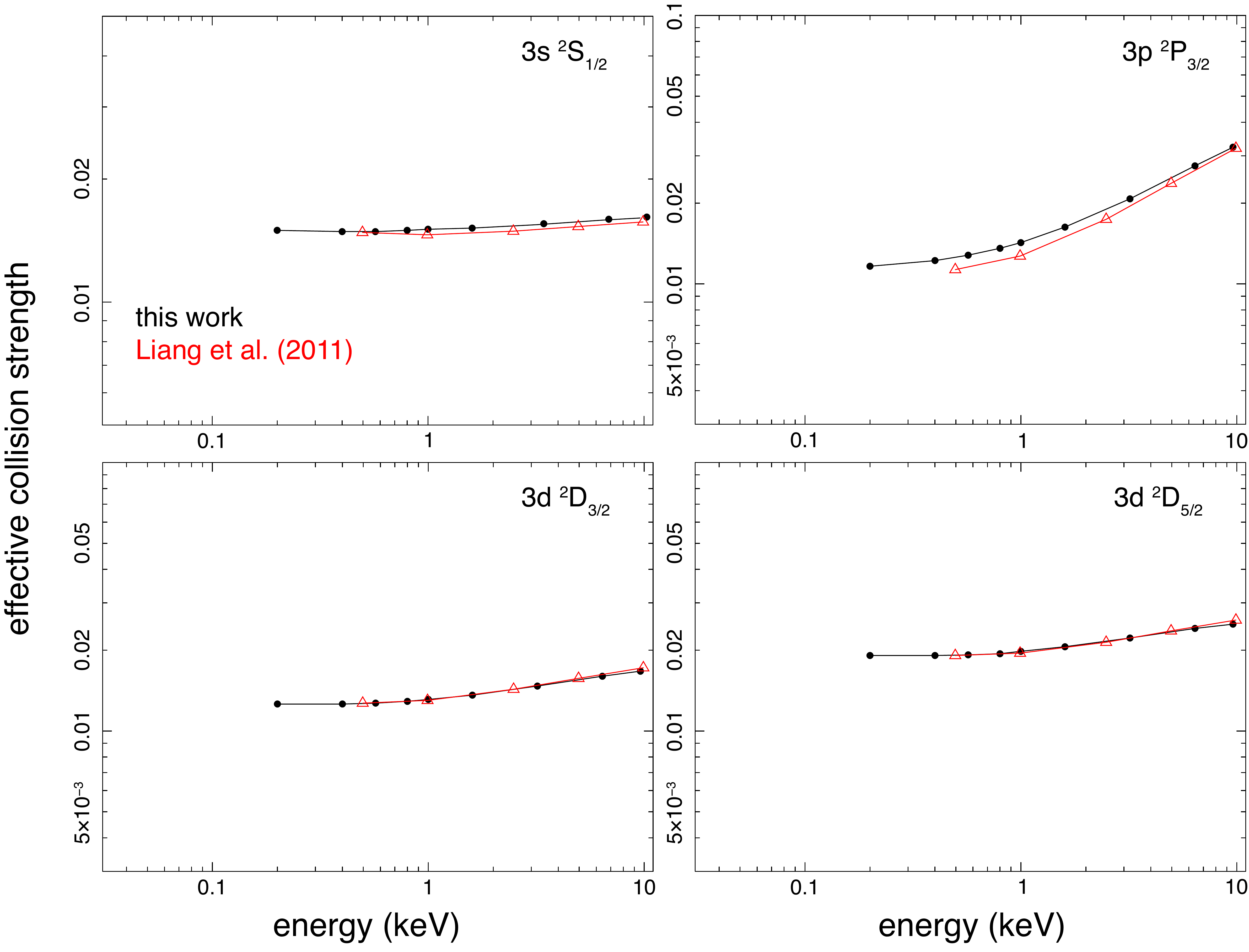}}
\caption{Same as Fig.~\ref{fig:rmatrix1} but for four \ion{Fe}{XXIV} lines. The $R$-matrix data from \citet{liang2011} are plot in red.}
\label{fig:rmatrix6}
\end{figure*}

\subsection{Comparing with recent $R$-matrix results: main transitions}
\label{sec:new_r_result}

More recently, new $R$-matrix calculations of Ne-like species were performed by \citet{loch2006} and \citet{liang2010}. The new calculations include more close-coupling expansions
than the earlier Breit-Pauli work by \citet{chen2002}. For \ion{Fe}{XVII}, the total atomic levels are 139 levels in \citet{loch2006} and 209 levels in \citet{liang2010}.
Here we further compare our calculation with the two new results for the four 3$s$ levels of \ion{Fe}{XVII}. The effective collision strengths are taken from
the OPEN-ADAS database\footnote{http://open.adas.ac.uk, ADF04}. 

As shown in Fig.~\ref{fig:rmatrix1}, the differences among the three datasets are reasonably small. Our calculation agrees with the $R$-matrix results within 
$\sim 5$\% for the $^3$P$_2$ level in $0.1-1.0$ keV, and within $5-20$\% for the $^3$P$_0$ level. For the $^1$P$_1$ and $^3$P$_1$ levels, our results are 
slightly lower, by $3-10$\% than those of \citet{liang2010} and $15-20$\% than those of \citet{loch2006}. In general, the differences observed between the FAC and 
the recent $R$-matrix calculations are well within the uncertainties among different $R$-matrix results. The agreement on the Ne-like lines is even better than
those on the H-like and He-like species as reported in \citet{atomic2017}. 

The resonance and direct excitation of $n=3$ and 4 for \ion{Fe}{XVIII} was calculated by \citet{witt2006}, using $R$-matrix with 195 levels. As shown in Fig.~\ref{fig:rmatrix2},
we compare the calculations of two 3$s$ levels which produce the strong 16~{\AA} line. The FAC and $R$-matrix results again agree within $10$\%, which might 
potentially be ascribed
to the uncertainties of the atomic structure for this nine-electron system. 

A similar $R$-matrix tool was used to calculate the electron collisional data for transitions among 342 levels of \ion{Fe}{XIX} \citep{butler2008}. As shown in 
Fig.~\ref{fig:rmatrix22}, we compare the calculations of four key excited levels with $n=3$. For the $3s$ level, the FAC value is higher than the $R$-matrix one by 
20\% at 0.2~keV, and by 13\% at 1~keV. As for the $3p$ and $3d$ levels, the two theoretical models seem to agree within 10\% in the temperature range of astrophysical interest. 

The electron collision strengths for a total of 302 close-coupling levels of \ion{Fe}{XX} were reported in \citet{witt2007}, based on the $R$-matrix theory using an 
intermediate-coupling frame transformation method. Their results are compared with the FAC calculation, for four low-lying levels shown in Fig.~\ref{fig:rmatrix23}. 
The comparison on the $3s$ $^4$P$_{5/2}$ level reveals a maximum discrepancy of $\sim 20$\% at 0.4~keV. For the other three levels, the difference between
the two datasets are reasonably small. 

A $R$-matrix calculation for C-like Fe excitation was reported by \citet{badnell2001}. It was carried out with the intermediate coupling frame transformation method 
for 200 close-coupling levels. In Fig.~\ref{fig:rmatrix3}, we show collision strengths for four representative levels, which are among the upper levels that produce 
the brightest emission lines. Our calculation suggests that these levels are significantly contributing to the resonant excitation. The FAC and $R$-matrix results are 
in accord well within $< 10$\% for $T_{\rm e} = 0.1-10$ keV. 

We further test the electron-impact excitation of \ion{Fe}{XXII} with the 204-level $R$-matrix results reported in \citet{liang2012}. As plotted in Fig.~\ref{fig:rmatrix4},
detailed comparisons for the effective collision strengths are made for four representative levels, which are selected again based on the related line emissivities and
the resonance contribution. For $2s^{2}2p$ $^2$P$_{3/2}$ and $2s^23s$ $^2$S$_{1/2}$, the two calculations in general agree for high temperatures, while the FAC values are
lower than the $R$-matrix values by $\sim 20$\% at $\leq 1$ keV. The two approaches converge well for the other two levels $2s2p^2$ $^2$P$_{1/2}$ and 
$2s^23d$ $^2$D$_{3/2}$.

The $R$-matrix calculation of electron-impact excitation of \ion{Fe}{XXIII} was made by \citet{fern2014}, based on 238 fine-structure levels in both the configuration interaction target and close-coupling collision expansions. We compare  
the collision strengths for four strong \ion{Fe}{XXIII} lines with their calculations, on
which the resonance has a large effect. As shown in Fig.~\ref{fig:rmatrix5}, the two data agree very well within a difference of $\leq 5$\% for $T_{\rm e} = 0.1-10$ keV.

In Fig.~\ref{fig:rmatrix6} we compare our calculations of the total excitation of four main \ion{Fe}{XXIV} lines with the $R$-matrix results reported 
in \citet{liang2011}. The $R$-matrix work was done with the intermediate coupling frame transformation method for 195 levels, and the Auger- as 
well as radiation- damping effects were both taken into account. For $3s$ $^2$S$_{1/2}$, $3d$ $^2$D$_{3/2}$, and $3d$ $^2$D$_{5/2}$, the $R$-matrix and our
data agree well with $\sim 5$\%, while for $3p$ $^2$P$_{3/2}$, the difference between the two approaches become slightly larger, $\sim 10$\%, for the
collision strengths at energy of $\leq 1$ keV.

The illustrative comparison indicates that the difference between the modern large-scale isolated resonance calculation and the recent 
interacting $R$-matrix becomes now smaller
than previously reported in \citet{badnell1994}. 
The two methods are well in line within $20$\% for the direct transitions
from the ground states to a few selected excited levels at $n=3$. 
As shown in Fig.~\ref{fig:17atot}, some excited levels are heavily populated 
through cascade, indicating that the line intensities could actually be significantly affected by the excitation rate coefficients 
to higher states. In the following section, we will give an extensive comparison for all the relevant rates.

%A portion of the uncertainties might come
%from the $R$-matrix side: as seen above, different $R$-matrix calculations are based on 
%different setups, while the calculation in the current work
%is done in a uniform way for all the iso-electronic sequences.

\subsection{Comparing with recent $R$-matrix results: scatter plots}
\label{sec:scatter_plot}

\begin{figure*}[!htbp]
\centering
\resizebox{0.85\hsize}{!}{\includegraphics[angle=0]{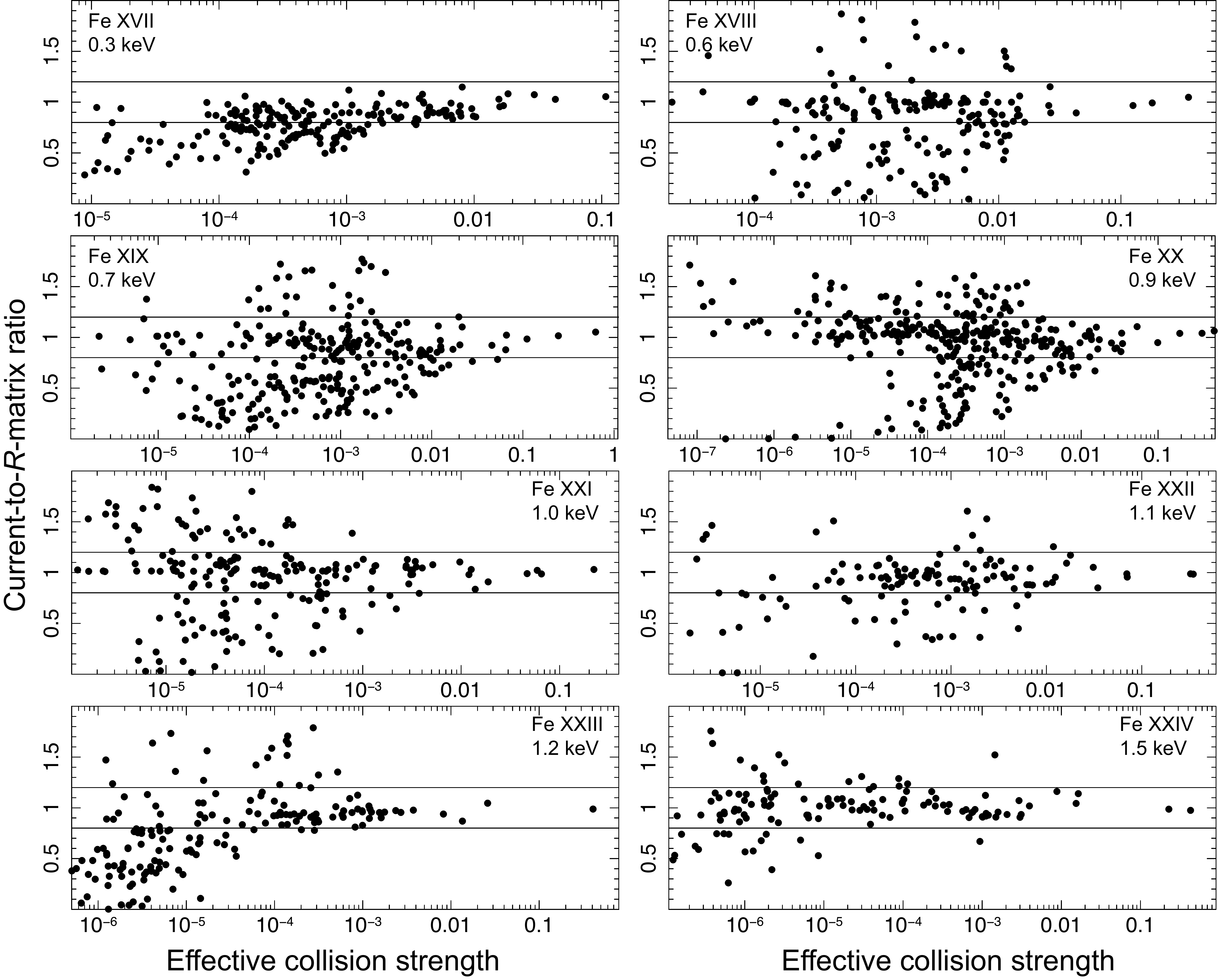}}
\caption{Ratios of the effective collision strengths from the current calculations to those from the recent $R$-matrix calculations, for
all excitations from the ground states of the Fe-L species. The comparison is made at the temperature with peak ionization concentration
of each ion (see Fig.~\ref{fig:facspec}). The horizontal lines indicate 20\% differences. }
\label{fig:scatter}
\end{figure*}

The scatter plot in Fig.~\ref{fig:scatter} shows the comparison of the effective 
collision strengths from the current and $R$-matrix calculations, for all the excitations
from the ground states. The $R$-matrix results are a collection of the works mentioned in
\S~\ref{sec:new_r_result}. The plot confirms that the two methods agree on the strongest transitions 
within uncertainties of $\sim 20$\%. However, the 
discrepancies between the two methods on the weaker transitions are substantially
larger, up to a factor of two for \ion{Fe}{XVII} and \ion{Fe}{XXIV}, and one order of magnitude for 
the rest ions, at the lowest effective collision strengths. Note that orders of magnitude difference for weak 
transitions are also found among different R-matrix calculations (ICFT, DARC, BSR, \citealt{fern2017} and references therein). 
These weak transitions would affect the satellite lines directly, and might also influence the main spectral lines
collectively through cascade. As shown in Fig.~\ref{fig:facspec}, the differences on the main transitions and on the weak transitions are clearly reflected
in the model spectra, indicating that both contribute systematic uncertainties that would 
significantly affect the astrophysical spectral measurements.

\end{appendix}

\end{document}